\documentclass[10pt]{article}
\usepackage{amssymb}
\usepackage{amsmath}
\usepackage{amsthm}
\usepackage{latexsym}
\usepackage[dvips]{epsfig}
\usepackage{mathrsfs}
\usepackage{bm}
\usepackage[utf8]{inputenc}
\usepackage[T1]{fontenc}
\usepackage{authblk}

\theoremstyle{plain}
\newtheorem{proposition}{Proposition}
\newtheorem{lemma}{Lemma}
\newtheorem{theorem}{Theorem}

\newtheorem*{main}{Theorem}

\newtheorem{remark}{Remark}

\font\SYM=msbm10
\newcommand{\Real}{\mbox{\SYM R}}

\newcommand{\Sphere}{\mbox{\SYM S}}

\font\tenscr=rsfs10 scaled1100
\font\sevenscr=rsfs7 
\font\fivescr=rsfs5 
\skewchar\tenscr='177
\skewchar\sevenscr='177
\skewchar\fivescr='177
\newfam\scrfam
\textfont\scrfam=\tenscr
\scriptfont\scrfam=\sevenscr
\scriptscriptfont\scrfam=\fivescr

\def\i{\iota}

\newcommand{\tensor}[3]{_{#1\phantom{#2}#3}^{\phantom{#1}#2}}
\newcommand{\half}{\frac{1}{2}}

\newcommand{\er}{{\underline{r}}}
\newcommand{\es}{{\underline{s}}}

\newcounter{mnote}

\begin{document}


\title{\textbf{A conformal approach for the analysis of the non-linear
    stability of pure radiation cosmologies}}

\author[,1,2]{Christian L\"ubbe \footnote{E-mail address:{\tt c.luebbe@qmul.ac.uk}, {\tt cl242@le.ac.uk}}}
\author[,2]{Juan Antonio Valiente Kroon \footnote{E-mail address:{\tt j.a.valiente-kroon@qmul.ac.uk}}}

\affil[1]{Department of Mathematics, University of Leicester, University Road, LE1 8RH, United Kingdom.}

\affil[2]{School of Mathematical Sciences, Queen Mary, University of London,
Mile End Road, London E1 4NS, United Kingdom.}

\maketitle

\begin{abstract}
The conformal Einstein equations for a tracefree (radiation) perfect fluid are
derived in terms of the Levi-Civita connection of a conformally rescaled metric.
These equations are used to provide a non-linear
stability result for de Sitter-like tracefree (radiation) perfect fluid
Friedman-Lema\^{\i}tre-Robertson-Walker cosmological models. The
solutions thus obtained exist globally towards the future and are
future geodesically complete.

\end{abstract}


\section{Introduction}
The conformal Einstein field equations have proven a powerful tool to
analyse the stability and the global properties of vacuum, electro-vacuum
and Yang-Mills-electro-vacuum spacetimes ---see
e.g. \cite{Fri86b,Fri91,Fri95,LueVal09a,LueVal10,LueVal11}. By
contrast, to the best of our knowledge, there has been no attempt to
make use of conformal methods to analyse similar issues in spacetimes
whose matter content is given by a perfect fluid. In this article we
make a first step in this direction. We discuss the stability and the 
global properties of a class of cosmological spacetimes having as a
source a perfect fluid with tracefree energy-momentum tensor.  The
solutions we construct are non-linear perturbations of a
Friedman-Lema\^{\i}tre-Robertson-Walker (FLRW) reference spacetime. 

\medskip
The present analysis is to be regarded as a first step in the
development of conformal methods for the discussion of cosmological
models whose matter content is described by a perfect fluid. Hence, we
restrict our attention to the simplest case from the point of view of
conformal methods: perturbations of a traceless prefect fluid
cosmological model with compact spatial sections of positive constant
curvature. Generalisation of our analysis to more general background
solutions 
and equations of state will be discussed elsewhere.

\medskip
The problem of the non-linear stability of FLRW cosmologies and the
exponential decay of perturbations is considered in
\cite{Reu99}. In that reference, a frame formulation of the Einstein-perfect fluid
system \cite{Fri98c} is used to obtain a suitable symmetric hyperbolic
evolution system for which the Kreiss-Lorenz theory can be readily
applied ---see \cite{KreLor98}. The results obtained hold for a large
class of equations of state, but not very stiff ones ---like the pure
radiation case discussed in the present article. More recently, the
problem of the non-linear stability of the irrotational Euler-Einstein
system for de Sitter-like spacetimes has been analysed in
\cite{RodSpe09}. This analysis shows that FLRW background solutions
with pressure $\tilde{p}$ and density $\tilde{\rho}$ related by a
barotropic equations of state of the form $\tilde{p}=\gamma
\tilde{\rho}$ with 
$1<\gamma <\tfrac{4}{3}$ are future asymptotically stable under small
irrotational perturbations. An extension of this analysis to the case
of fluids with non-zero vorticity has been given in \cite{Spe11}.

\medskip
It is notable that the case of a pure radiation perfect fluid cannot
be covered by the analysis of \cite{Reu99,RodSpe09,Spe11}. By
contrast, from the point of view of conformal methods, the pure
radiation perfect fluid case turns out to be one of the simplest
scenarios to be considered. Finally,  it should be mentioned that conformal
methods have been used to pose an initial value problem for the
Einstein-Euler system at the Big Bang for a class of
cosmological models with isotropic singularities ---see
\cite{AngTod99a}. The methods used in that work do not allow, however,
to obtain global existence assertions towards the future. 

\medskip
Our main result can be stated as
follows:

\begin{main}
Suppose one is given Cauchy initial data for the Einstein-Euler system
with a de Sitter-like cosmological constant and equation of state for
pure radiation. If the initial data is sufficiently close to data for a FLRW
cosmological model with the same equation of state, value of the
cosmological constant and spatial curvature $k=1$, then the development
exists globally towards the future, is future geodesically complete
and remains close to the FLRW solution.
\end{main}

A detailed and technically precise version of this result is given in
Theorem \ref{MainTheorem}.

\begin{remark}
{\em  Similar future global existence and stability results
can be obtained using the methods of this article for a FLRW
background solution with pure radiation equation of state, de
Sitter-like or vanishing cosmological constant, $\lambda$, and
$k=0,\,-1$. These models expand indefinitely towards the future, and
remarkably, their scale factor can be computed explicitly ---see
\cite{GriPod09}. In the cases with $\lambda=0$, minor technical
modifications need to be introduced to account for a null conformal
boundary. The stability of these models will be discussed elsewhere
by means of different (conformal) methods.}
\end{remark}

\begin{remark}
{\em The restriction of our analysis to the case of perfect fluids
  with traceless energy-momentum tensor is technical: in this case the equation of
  conservation of energy momentum transforms homogeneously under
  conformal transformations. In the case of perfect fluids with an
  energy-momentum tensor with non-vanishing trace a regularisation of
  the rescaled equations of motions must be carried out. The analysis
  for the wave equation in \cite{BicSchTod10b,Hub95} may be a guide for this type of
  generalisation of our analysis.}
\end{remark}

\subsection*{Structure of the article}
The article is organised as follows: Section
\ref{Section:NotationConventions} provides a summary of the tensorial
conventions to be used in the present article. Furthermore, in
Subsection \ref{Subsection:Coordinates3Sphere} a discussion of the
procedure of how to coordinatise and introduce frame fields of the 3-sphere,
$\Sphere^3$ is provided.  Section \ref{Section:gneralRemarks} provides
general remarks concerning perfect fluid cosmological models and a
summary of the properties of the background solutions required in our
subsequent analysis. These are summarised in Proposition
\ref{Proposition:ScaleFactor}. Section \ref{Section:XCFE} gives a
brief summary of the conformal Einstein field equations with
matter. Section \ref{Section:PerfectFluids} provides a discussion of
the Euler equations in the context of the conformal field
equations. In Section \ref{Section:derive_SHsystem} we  
discuss gauge considerations and the procedure leading to a
hyperbolic reduction of the conformal field equations. The keys steps
in this procedure have been discussed extensively elsewhere, so that
this discussion is kept to a minimum. In particular, Subsection
\ref{Subsection:SH_system} provides a summary of the structural
properties of the conformal evolution equations while Subsection
\ref{Section:PropagationConstraints} analyses the issue of the
propagation of the constraints. Section \ref{FLRW:CFE} casts the FLRW
background  as a solution of the conformal field equations of
Section \ref{Section:XCFE}, and analyses some of its
properties. Finally, Section \ref{Section:Existence} is concerned with
our main result ---the existence and stability result for perfect
fluid cosmologies with a de Sitter-like cosmological constant as given in
Theorem \ref{MainTheorem}.

\section{Notation and conventions}
\label{Section:NotationConventions}

\subsection{Index and curvature conventions}
Throughout this article we work with a spacetime
$(\tilde{\mathcal{M}},\tilde{g}_{\mu\nu})$, where
$\tilde{g}_{\mu\nu}$, ($\mu,\nu=0,1,2,3$) is a Lorentzian metric with
signature $(+,-,-,-)$. We will denote by $\tilde{\nabla}$ the
Levi-Civita connection of $\tilde{g}_{\mu\nu}$ ---that is, the unique
torsion-free connection that preserves the metric
$\tilde{g}_{\mu\nu}$. In the sequel, 
$\tilde{R}_{\mu\nu\lambda\rho}$, $\tilde{R}_{\mu\nu}$ and $\tilde{R}$
will denote, respectively, the Riemann curvature tensor, the Ricci tensor
and the Ricci scalar of the Levi-Civita connection
$\tilde{\nabla}$. The conventions for the curvature used in this
article are such that
\begin{equation}
\label{Riemann:convention}
\tilde{R}^\mu{}_{\nu\lambda\rho} \xi^\nu =\left(\tilde{\nabla}_\lambda \tilde{\nabla}_\rho -\tilde{\nabla}_\rho \tilde{\nabla}_\lambda \right)\xi^\mu, \quad\quad
\tilde{R}_{\mu\nu}=\tilde{R}^\lambda{}_{\nu\lambda\mu}, \quad\quad
\tilde{R} = \tilde{R}_{\mu\nu} \tilde{g}^{\mu\nu} .
\end{equation}
As a consequence of our signature conventions, then $\lambda<0$
corresponds to de Sitter-like values of the cosmological constant,
while $\lambda>0$ corresponds to anti-de Sitter-like values. While
$\mu,\;\nu,\ldots$ denote spacetime tensorial indices,
$\alpha,\;\beta,\ldots$ denote spatial tensorial ones. Most of our
discussion will be based on a frame formalism in which $i,\;j,\ldots$
denote spacetime indices ranging $0,\ldots,3$. Similarly,
$a,\;b,\ldots$ will denote spatial indices ranging
$1,\;2,\;3$. Spinorial expressions and arguments will be used
routinely, and we will follow the conventions of
\cite{PenRin84}. Consequently, the indices $A,\; B,\ldots$ will
be spinorial ones.

\subsection{Coordinates and vector fields on the 3-sphere}
\label{Subsection:Coordinates3Sphere}
The present analysis will be concerned with spacetimes which are
conformal to manifolds with topology $I\times \Sphere^3$ where $I$ is
an open interval on $\Real$. In what follows, the manifold $\Sphere^3$ will
always be thought of as the following submanifold of $\Real^4$:
\[
\Sphere^3 =\left \{ x^\mathcal{A} \in \Real^4 \;\bigg | \; (x^1)^2 + (x^2)^2 + (x^3)^2 + (x^4)^2=1  \right\}.
\]
The restrictions of the functions $x^\mathcal{A}$,
$\mathcal{A}=1,2,3,4$ on $\Real^4$ to $\Sphere^3$ will again be
denoted by $x^\mathcal{A}$. The vector fields
\begin{subequations}
\begin{eqnarray}
&& c_1 \equiv x^1 \partial_4 -x^4\partial_1 +x^2\partial_3 -x^3\partial_2, \label{v_field_c1} \\
&& c_2 \equiv x^1 \partial_3 -x^3\partial_1 +x^4\partial_2 -x^2\partial_4, \label{v_field_c2}\\
&& c_3 \equiv x^1 \partial_2 -x^2\partial_1 +x^3\partial_4 -x^4\partial_3, \label{v_field_c3}
\end{eqnarray}
\end{subequations}
on $\Real^4$ are tangent to $\Sphere^3$. In the sequel, they will
always be considered as vectors on $\Sphere^3$. The vector fields
$\{c_\er\}\equiv\{c_1, \,c_2,\,c_3\}$ constitute a globally defined
frame on $\Sphere^3$ which is orthonormal with respect to the standard
metric of $\Sphere^3$. Moreover, the frame $\{c_1, \,c_2,\,c_3\}$ can
be completed with a vector $c_0$ which is orthonormal to the standard
metric on $I\times \Sphere^3$, $\{c_\es\}\equiv \{ c_0,\,c_1,
\,c_2,\,c_3 \}$.

\medskip
Let $(\mathcal{M},g_{\mu\nu})$ be a spacetime such that the manifold
$\mathcal{M}$ is diffeomorphic to $\Real\times \Sphere^3$. A map
$\Phi$ defined on an open subset $\mathcal{U}\subset \mathcal{M}$ will
be said to be a \emph{cylinder map} if it maps $\mathcal{U}$
diffeomorphically onto a set $I\times \Sphere^3$, such that the sets
$\Phi^{-1}(\{\tau\}\times \Sphere^3)$ are spacelike Cauchy
hypersurfaces of $\mathcal{M}$ and the curves $I\ni \tau \rightarrow
\Phi^{-1}(\tau,p)\subset \mathcal{M}$, $p\in \Sphere^3$ are timelike
with respect to the metric $g_{\mu\nu}$. The cylinder map will be used
to pull-back to $\mathcal{U}$ the coordinates
$(\tau,x^\mathcal{A})\equiv(\tau,{\bm x})$ in $I \times
\Sphere^3$. Furthermore, one can use $\Phi$ to pull-back to
$\mathcal{U}$ the frame fields $c_\es$ defined in the previous
paragraph. For simplicity of notation, such pull-back will be denoted
again by $c_\es$.

\section{General remarks about FLRW cosmological models}
\label{Section:gneralRemarks}

A cosmological model
$(\tilde{\mathcal{M}},\tilde{g}_{\mu\nu},\tilde{u}^\mu)$ is a
representation of the universe at a particular averaging scale. It is
defined by a Lorentzian
metric $\tilde{g}_{\mu\nu}$ on the manifold
$\tilde{\mathcal{M}}$ and by a family of fundamental observers whose
congruence of worldlines is represented by the timelike 4-velocity
$\tilde{u}^\mu$ ---usually taken to be the matter 4-velocity. 
It is usually assumed that this congruence is expanding at some time. 
These assumptions together with a specification of the matter content are
used to determine the dynamics of the universe. In what follows, it
will be assumed that the interaction between geometry and matter is
described by the Einstein field equations
\begin{equation}
\label{EFEMatter}
\tilde{R}_{\mu\nu} -\tfrac{1}{2} \tilde{R} \tilde{g}_{\mu\nu} +\lambda
\tilde{g}_{\mu\nu}= \tilde{T}_{\mu\nu},
\end{equation}
and the energy-momentum conservation equation
\begin{equation}
\tilde{\nabla}^\mu \tilde{T}_{\mu\nu}=0.
\label{PhysicalEnergyMomentumConservation}
\end{equation}
As already mentioned,  the conventions for the cosmological constant $\lambda$ used in the
present article are such that in vacuum, the case $\lambda<0$
describes a de Sitter-like spacetime, while the case $\lambda>0$
corresponds to an anti-de Sitter-like one. 

\medskip
Our discussion will be concerned with energy-momentum tensors of perfect
fluids for which
\[
\tilde{T}_{\mu\nu} = (\tilde{\rho}+\tilde{p}) \tilde{u}_\mu \tilde{u}_\nu - \tilde{p} \tilde{g}_{\mu\nu},
\]
where $\tilde{\rho}$, $\tilde{p}$ and $\tilde{u}^\mu$ denote,
respectively, the density, pressure and 4-velocity of the cosmological
fluid. The fluid 4-velocity $\tilde{u}^\mu$ is timelike and satisfies
the normalisation condition $\tilde{u}_\mu \tilde{u}^\mu=1$.

\medskip
The background solution whose non-linear stability will be considered
in the present article belongs to the family of so-called
Friedman-Lema\^{\i}tre-Robertson-Walker (FLRW) cosmological models. The
FLRW models are homogeneous and isotropic. Their line element is
usually given in the form
\begin{equation}
\tilde{g}_{\mathscr{F}} \equiv \tilde{g}_{\mu\nu} \mbox{d} x^\mu \mbox{d}x^\nu= \mbox{d}t^2 - \frac{a^2(t)}{\left(1+\tfrac{1}{4}kr^2 \right)^2}\left(\mbox{d}r^2 +r^2 \mbox{d}\theta^2 +r^2 \sin^2\theta \mbox{d}\varphi^2 \right),
\label{FRW:LineElement}
\end{equation}
where $a(t)$ is the so-called \emph{scale factor}. This metric
automatically defines a perfect fluid energy-momentum tensor. When
$k=0$ the spatial sections are flat, if $k<0$ the spatial sections
have negative curvature, while if $k>0$, the spatial sections have
positive curvature.\emph{ The present analysis is concerned with FLRW
cosmologies with spatial sections of positive curvature ($k=1$)} for
which coordinates can be introduced such that:
\begin{equation}
\tilde{g}_{\mathscr{F}} = \mbox{d}t^2 -a^2(t)\mbox{d}\sigma^2,
\label{FRW:AltLineElement}
\end{equation}
with
\[
\mbox{d}\sigma^2\equiv \mbox{d}\psi^2 + \sin^2\psi \mbox{d}\theta^2 + \sin^2\psi \sin^2\theta \mbox{d}\varphi^2,
\]
the standard line element of $\Sphere^3$ in polar coordinates. If
the cosmological fluid satisfies the barotropic equation of state
$\tilde{p} =(\gamma -1) \tilde{\rho}$, where $1\leq \gamma \leq 2$ is
a constant, then the evolution of the scale factor $a(t)$ is governed
by the \emph{Friedmann equation}:
\begin{equation}
\frac{\dot{a}^2}{a^2} = -\tfrac{1}{3}\lambda - \frac{1}{a^2} + \frac{c}{a^{3\gamma}},
\label{FriedmannEquation}
\end{equation}
where $c$ is a constant. \emph{In what follows we will only be
concerned with the case $\gamma=\tfrac{4}{3}$ corresponding to the so-called
traceless perfect fluid (pure radiation). Furthermore, we
assume $\lambda<0$.} Equation \eqref{FriedmannEquation} admits a static
(i.e. time independent solution) in which the values of the scale
factor and the cosmological constant are related by:
\begin{equation}
a(t)=a_0=\mbox{constant}, \qquad \lambda=\lambda_0\equiv \tfrac{3}{2} a_0^{-2}. \label{StaticFLRW}
\end{equation}
In the dynamical case, under the assumptions $\gamma=\tfrac{4}{3}$, $\lambda<0$,
the Friedmann equation \eqref{FriedmannEquation} can be integrated
explicitly ---see e.g. \cite{GriPod09}. Different types of solutions
are obtained, depending on the relative value of $\lambda$ with
respect to $\lambda_0$ as given in equation \eqref{StaticFLRW}, where
$a_0\neq0$ is now the value of the scale factor at some fiduciary time
$t=t_0\neq0$. The relevant properties for the analysis of these solutions
are summarised in the following proposition:

\begin{proposition}
\label{Proposition:ScaleFactor}
For a FLRW cosmology with $k=1$, $\gamma=\tfrac{4}{3}$ and $\lambda<0$,
$\lambda\neq\lambda_0$, the scale factor, $a(t)$, is a smooth,
non-vanishing and monotonically increasing function for
$t\in[t_0,\infty)$, with $t=t_0>0$ and $a_0=a(t_0)>0$. Furthermore,
\[
\int^\infty_{t_0} \frac{\mbox{d}s}{a(s)}<\infty, 
\]
and one has the limits
\[
a\rightarrow \infty, \quad \dot{a}/a\rightarrow \sqrt{-\tfrac{1}{3}\lambda}, \quad \ddot{a}/a\rightarrow -\tfrac{1}{3}\lambda.
\]
as $t\rightarrow \infty$. The pressure for these models is given by
\[
\tilde{\rho}= \tilde{\rho}_0 a_0^4/a^4,
\]
where $\tilde{\rho}_0=\tilde{\rho}(t_0)$. In particular, one has that
$\tilde{\rho}\rightarrow 0$ as $t\rightarrow \infty$.
\end{proposition}

The proof of this proposition follows from direct inspection of the
explicit solutions ---see e.g. \cite{GriPod09}, page 78.

\begin{remark}
{\em A similar type of result can be obtained for FLRW
models with $\gamma=\tfrac{4}{3}$, $\lambda\leq 0$ and $k=-1,
\,0$. Again, see \cite{GriPod09}.} 
\end{remark}

\section{The conformal field equations with matter}
\label{Section:XCFE}

The stability of the solutions to the Einstein equations described by
the metric $\tilde{g}_{\mu\nu}$ corresponding to the line element
\eqref{FRW:LineElement} will be analysed in terms of a conformally
related (\emph{unphysical)} metric $g_{\mu\nu}$. This strategy leads
to consider the conformal Einstein field equations. The idea of
\emph{vacuum conformal Einstein field equations} expressed in terms of
the Levi-Civita connection $\nabla$ of the metric $g_{\mu\nu}$ and
associated objects was originally introduced in
\cite{Fri81b,Fri81a,Fri83}. The generalisation of these conformal
equations to physical spacetimes containing matter was discussed in
\cite{Fri91}. More recently, a more general type of vacuum conformal
equations ---the \emph{extended conformal Einstein field equations}---
expressed in terms of a Weyl connection $\hat{\nabla}$ has been
introduced ---see \cite{Fri95}.

\subsection{Conformal rescalings}
All throughout we assume that the two metrics $\tilde{g}_{\mu\nu}$
and $g_{\mu\nu}$ are conformally related to each other via
\begin{equation}
g_{\mu\nu} = \Theta^2 \tilde{g}_{\mu\nu}, \label{ConformalRescalingMetric}
\end{equation}
where $\Theta$ is a non-negative scalar field ---the conformal
factor. The Christoffel symbols $\tilde{\Gamma}_\mu{}^\rho{}_\nu$ and
$ \Gamma_\mu{}^\rho{}_\nu$ of the associated Levi-Civita connections
$\tilde{\nabla}$ and $\nabla$ are related by
\begin{equation}
\tilde{\Gamma}_\mu{}^\rho{}_\nu - \Gamma_\mu{}^\rho{}_\nu = S_{\mu\nu}{}^{\rho\lambda}\Upsilon_\lambda,
\label{LCConnection:TransformationRuleConnection}
\end{equation}
where $\Upsilon_\lambda = \Theta^{-1} \nabla_\lambda \Theta$ and
$S_{\mu\nu}{}^{\rho\lambda}$ is the conformally invariant tensor
\[
S_{\mu\nu}{}^{\lambda\rho} = \delta_\mu{}^\lambda \delta_\mu{}^\rho +
\delta_\mu{}^\rho \delta_\nu{}^\lambda - g_{\mu\nu} g^{\lambda\rho}. 
\]

\subsection{Curvature tensors}

In a 4-dimensional spacetime the \emph{Schouten tensor},
${P}_{\mu\nu}$, of the connection $\nabla$ is defined by
\[
{P}_{\mu\nu} = \tfrac{1}{2} R_{\mu\nu} - \tfrac{1}{12} R g_{\mu\nu}.
\]
The Schouten tensor of the connection $\tilde{\nabla}$ is defined by a
similar expression involving the physical Ricci tensor and scalar. The
tensors $\tilde{P}_{\mu\nu}$ and $P_{\mu\nu}$ are related by
\begin{equation}
\label{transform_Schouten}
P_{\mu\nu}-\tilde{P}_{\mu\nu}=\nabla_\mu \Upsilon_\nu - \Upsilon_\mu \Upsilon_\nu +\half g_{\mu\nu} \Upsilon_\rho \Upsilon^\rho .
\end{equation}
We can thus decompose the Riemann curvature tensor,
$R^\mu{}_{\nu\lambda\rho}$, of the connection $\nabla$ into its irreducible parts as 
\begin{eqnarray}
&& {R}^\mu{}_{\nu\lambda\rho} = C^\mu{}_{\nu\lambda\rho} + 2
S_{\nu[\lambda}{}^{\mu \sigma}{P}_{\rho] \sigma},  \nonumber \\
&& \phantom{{R}^\mu{}_{\nu\lambda\rho}} = C^\mu{}_{\nu\lambda\rho} +2 \left( g^\mu{}_{[\lambda} {P}_{\rho]\nu}
 -  g_{\nu[\lambda}{P}_{\rho]}{}^\mu \right), \label{Curvature:IrreducibleDecomposition}
\end{eqnarray}
where $C^\mu{}_{\nu\lambda\rho}$ denotes the conformally invariant \emph{Weyl tensor}.

\medskip
As $\nabla$ is a Levi-Civita connection it satisfies the
\emph{first} and \emph{second Bianchi identities}:
\begin{subequations}
\begin{eqnarray}
&& {R}^\mu{}_{[\nu\lambda\rho]}=0, \label{WeylConnection:FirstBianchiIdentity} \\
&& {\nabla}_{[\sigma} {R}^\mu{}_{|\nu|\lambda\rho]}=0.  \label{WeylConnection:SecondBianchiIdentity}
\end{eqnarray}
\end{subequations}

In our discussion of the conformal field equations with matter we will
make use of the physical and unphysical Cotton-York tensors
$\tilde {Y}_{\lambda\rho\nu}$ and ${Y}_{\lambda\rho\nu}$ given, respectively, by
\[
\tilde{Y}_{\lambda\rho\nu} \equiv 
\tilde{\nabla}_\lambda \tilde{P}_{\rho\nu} - \tilde{\nabla}_\rho \tilde{P}_{\lambda\nu},
\qquad 
Y_{\lambda\rho\nu} \equiv 
{\nabla}_\lambda {P}_{\rho\nu} - {\nabla}_\rho {P}_{\lambda\nu}.
\]
The tensor $Y_{\lambda\rho\nu}$  appears in the once contracted Bianchi identity
\begin{equation}
\label{WeylConnection:contractedBianchiIdentity}
{\nabla}_{\mu } {C}^\mu{}_{\nu\lambda\rho} = {Y}_{\lambda\rho\nu}.
\end{equation}
Finally, it is noticed that the twice contracted Bianchi identity takes the form
\begin{equation}
\label{WeylConnection:contractedBianchiIdentity}
{\nabla}^\nu {P}_{\rho\nu} =  {\nabla}_\rho {P},
\end{equation}
where ${P}=g^{\lambda \nu}{P}_{\lambda\nu} $. 

\subsection{Frame and spinor formulations}

In what follows, consider a frame field $\{e_i\}$, $i=0,\ldots,3$
which is orthogonal with respect to the metric $g_{\mu\nu}$. By
construction one has that
\begin{equation}
g_{\mu\nu} e_i{}^\mu e_j{}^\nu =\eta_{ij}, \qquad \eta_{ij}\equiv \mbox{diag}(1,-1,-1,-1).
\label{frame:metric}
\end{equation}

\medskip
 In order to discuss the extended conformal
Einstein field equations, it will be convenient to 
regard,  for the moment,  the connection $\nabla$ only as a metric connection with
respect to $g_{\mu\nu}$ ---i.e. $\nabla_\lambda g_{\mu\nu}=0$.
Under this assumption, the connection $\nabla$ could have torsion, and thus, it would not be a
Levi-Civita connection. The connection coefficients, $\Gamma_i{}^k{}_j$,
of $\nabla$ with respect to the frame $e_k$ are defined by the
relation
\[
\nabla_i e_j = \Gamma_i{}^k{}_j e_k.
\]
As a consequence of having a metric connection, the connection
coefficients satisfy
\[
\Gamma_i{}^k{}_j \eta_{kl} + \Gamma_i{}^k{}_l \eta_{kj}=0.
\]
The torsion, $\Sigma_i{}^k{}_j$, of the connection $\nabla$ is defined by
\[
\Sigma_i{}^k{}_j e_k \equiv \left(\Gamma_i{}^k{}_j - \Gamma_j{}^k{}_i
\right)e_k -[e_i,e_j].
\]
If $\Sigma_i{}^k{}_j=0$ so that the connection $\nabla$ is the unique
Levi-Civita connection of $g_{\mu\nu}$, the connection coefficients
acquire the additional symmetry 
\[
\Gamma_i{}^k{}_j = \Gamma_j{}^k{}_i.
\]

Related to the $g$-orthonormal frame $e_k$ we will consider a
normalised spinor dyad $\{ \delta_A \}$, $A=0,\,1$, such that 
\[
e_{AA'} = e_k \sigma^k{}_{AA'}
\]
where $\sigma^k{}_{AA'}$ are the constant van der Waerden symbols.

\medskip
In the sequel, a space spinor formalism will be introduced ---see
e.g. \cite{Som80}. To this
end, we consider a timelike spinor $\tau_{AA'}$ which in terms of the
dyad $\{\delta_A\}$ can be expressed as
\[
\tau^{AA'} = \epsilon_0{}^A \bar{\epsilon}_{0'}{}^{A'} +
\epsilon_1{}^A  \bar{\epsilon}_{1'}{}^{A'} .
\] 
In particular, one has the normalisation condition
$\tau_{AA'}\tau^{AA'}=2$. The space spinor formalism allows to turn
primed indices in spinorial expressions into unprimed ones by suitable
contractions with $\tau_A{}^{A'}$ ---see
\cite{Fri91,LueVal09a,LueVal10,LueVal11} for more details. We  simply
recall that the space spinor decomposition of a spinor $u_{AA'}$ is
given by
\begin{equation}
u_{AA'} = \tfrac{1}{2} u \tau_{AA'} - \tau^Q{}_{A'} u_{QA}, 
\label{u_spacespinors}
\end{equation}
where
\[
u\equiv u_{PP'} \tau^{PP'}, \qquad u_{AB}\equiv \tau^{P'}{}_{(B} u_{A)P'}.
\]

\subsection{The conformal field equations with tracefree matter}

In our subsequent discussion it will be convenient to distinguish
between the \emph{geometric curvature} ${r}^k{}_{lij}$ ---i.e. the
expression of the curvature related to the
connection coefficients ${\Gamma}_i{}^j{}_k$--- and the
\emph{algebraic curvature} ${R}^k{}_{lij}$ ---i.e. the
decomposition of the curvature in terms of irreducible components
given by equation \eqref{Curvature:IrreducibleDecomposition}. One
has that
\begin{eqnarray*}
&& {r}^k{}_{lij} \equiv e_i \left( {\Gamma}\tensor{j}{k}{l} \right) - e_j \left( {\Gamma}\tensor{i}{k}{l} \right) 
- {\Gamma}\tensor{m}{k}{l} \left( {\Gamma}\tensor{i}{m}{j} - {\Gamma}\tensor{j}{m}{i}\right) 
+ {\Gamma}\tensor{i}{k}{m}  {\Gamma}\tensor{j}{m}{l} 
- {\Gamma}\tensor{j}{k}{m}  {\Gamma}\tensor{i}{m}{l} + {\Sigma}_i{}^m{}_j {\Gamma}_m{}^k{}_l,  \\
&& {R}^k{}_{lij} \equiv C^k{}_{lij} +2 \left( \delta^k{}_{[i} {P}_{j]l}  
-  \eta_{l[i}{P}_{j]}{}^k \right)
 = C^k{}_{lij} + 2 S_{l[i}{}^{km}{P}_{j]m}.
\end{eqnarray*}

Following \cite{Fri03a}, in the sequel it will be convenient to introduce the
variables
\begin{subequations}
\begin{eqnarray}
&& d^k{}_{lij}\equiv \Theta^{-1}C^k{}_{lij}, \label{defining dklij}\\ 
&& d_i \equiv  \nabla_i \Theta  \label{defining di}, \\
&& {s} \equiv \tfrac{1}{4}({\nabla}^k d_k  + \Theta {P}^k{}_k) .
\label{defining_s}
\end{eqnarray}
\end{subequations}

\medskip
Furthermore, we also consider the  following
\emph{zero quantities} ---cfr. \cite{Fri91}:
\begin{subequations}
\begin{eqnarray}
&&\Sigma_i{}^l{}_j e_l\equiv \left( {\Gamma}\tensor{i}{l}{j} - {\Gamma}\tensor{j}{l}{i}\right) e_l -[e_i,e_j],\label{torsionfree}\\
&& \Xi^k{}_{lij}\equiv {r}^k{}_{lij} -{R}^k{}_{lij},\label{Curvature}\\
&& \Delta_{lij} \equiv {\nabla}_{i}{P}_{jl} - {\nabla}_{j}{P}_{il} -d_k d^k{}_{lij} -
\Theta^2 T_{ijl}, \label{Cotton-York} \\
&& \Lambda_{lij} \equiv \nabla_k d^k{}_{lij}  - \Theta T_{ijl}, \label{Bianchi} \\
&& \delta_k \equiv d_k - {\nabla}_k \Theta, \label{1-form} \\
&& \delta_{ij} \equiv {\nabla}_i d_j + \Theta {P}_{ij} - {s} \eta_{ij} - \tfrac{1}{2} \Theta^3 T_{ij}, \label{nabla_d} \\
&& \zeta_k \equiv {\nabla}_k {s} + d^l P_{kl} - \tfrac{1}{2} \Theta^2 d^l T_{lk},
\label{nabla_s}  \\
&& \zeta \equiv \lambda -6\Theta s + 3 d_k d^k, \label{Equation:lambda}
\end{eqnarray}
\end{subequations}
where 
\[
T_{kl} \equiv \Theta^{-2}\tilde{T}_{kl}, \qquad T_{ijk}\equiv\Theta^{-2} \tilde{Y}_{ijk} .
\]

\medskip
The interpretation of the zero quantities
\eqref{torsionfree}-\eqref{Bianchi} is as follows: the zero quantity
given by \eqref{torsionfree} measures the torsion of the connection
${\nabla}$; that of \eqref{Curvature} relates the expression of
the curvature of ${\nabla}$ with its decomposition in terms of
irreducible components. Equations \eqref{Cotton-York}
and \eqref{Bianchi} measure the deviation from the fulfillment of the
once contracted
Bianchi identity.  Finally, equations \eqref{1-form}, \eqref{nabla_d}
and \eqref{nabla_s} bring into play the definitions \eqref{defining
  di} and \eqref{defining_s} and give rise to differential conditions
for the fields $\Theta$, $d_i$ and $s$.  

\medskip
The \emph{conformal Einstein field equations with matter} are
then given by
\begin{subequations}
\begin{eqnarray}
&&  \Sigma_i{}^k{}_j e_k=0, \qquad \Xi^k{}_{lij}=0, \qquad \Delta_{lij}=0,
\qquad \Lambda_{lij}=0, \label{XCFEFrame1}\\
&& \qquad \delta_k=0,  \qquad \delta_{ij}=0,
\qquad \zeta_k=0, \qquad \zeta=0. \label{XCFEFrame2}
\end{eqnarray}
\end{subequations}

These equations yield differential conditions for the frame
coefficients $e_i$, the spin coefficients $\Gamma_i{}^j{}_k$, the components of the Schouten
tensor ${P}_{ij}$, the rescaled Weyl tensor $d^k{}_{lij}$,
the conformal factor $\Theta$, the 1-form $d_i$, and the
scalar $s$, respectively. As discussed in e.g. \cite{Fri03a}, equation
\eqref{Equation:lambda} has the role of a constraint which holds by
virtue of the other conformal field equations if it is satisfied on
some initial hypersurface.  It is noticed that as the torsion, ${\Sigma}_i{}^k{}_j$, is being
introduced as a zero quantity, it can be consistently set to zero in
the geometric curvature appearing in the definition for the zero
quantity ${\Xi}^k{}_{lij}$ ---equation \eqref{Curvature}.  

\medskip
Equations \eqref{XCFEFrame1}-\eqref{XCFEFrame2} need to be complemented with the
energy-momentum conservation equation
\eqref{PhysicalEnergyMomentumConservation}. Its particular details will depend on the matter model under
consideration.

\begin{remark}
{\em Using a direct generalisation of the arguments
presented in \cite{Fri81a,Fri81b} one can show that a solution to the
conformal Einstein field equations with matter
\eqref{XCFEFrame1}-\eqref{XCFEFrame2} and
\eqref{PhysicalEnergyMomentumConservation} give rise to a solution to
the physical Einstein-matter system
\eqref{EFEMatter}-\eqref{PhysicalEnergyMomentumConservation} ---see
also Theorem 3.1 in \cite{Fri83}.}
\end{remark}

\begin{remark}
{\em As a result of the conformal rescaling
\eqref{ConformalRescalingMetric}, the conformal equations
\eqref{XCFEFrame1}-\eqref{XCFEFrame2} have a built-in conformal
freedom which needs to be specified in order to deduce suitable evolution
equations for the conformal fields. Further gauge freedom in
equations \eqref{XCFEFrame1}-\eqref{XCFEFrame2} is concerned with the
partial specification of the frame $e_k$ and the choice of
coordinates. These will be specified by the choice of suitable gauge
source functions.} 
\end{remark}

\section{Perfect fluids in the context of the conformal approach}
\label{Section:PerfectFluids}

In this section we present a discussion of the relativistic equations
describing a perfect fluid which is geared towards our particular
applications. 

\subsection{The energy-momentum tensor and its transformation rules}
Given the spacetime
$(\tilde{\mathcal{M}},\tilde{g}_{\mu\nu})$, the energy-momentum tensor for a perfect fluid
with 4-velocity $\tilde{u}^i$, pressure $\tilde{p}$, and density $\tilde{\rho}$  has the form
\begin{equation}
\tilde{T}_{\mu\nu} = (\tilde{\rho} + \tilde{p} )\tilde{u}_\mu \tilde{u}_\nu -  \tilde{p} \tilde{g}_{\mu\nu}.
\label{ConformalTransformation:MatterFields}
\end{equation}
In order to perform a discussion of the perfect fluid in the
conformally rescaled (unphysical) spacetime one introduces unphysical
versions of the physical fields. More precisely, one defines
\[
T_{\mu\nu} \equiv \Theta^{-2}\tilde{T}_{\mu\nu}, \qquad u_\mu \equiv
\Theta \tilde{u}_\mu, \qquad
\rho\equiv \Theta^{-4}\tilde{\rho}, \qquad p\equiv\Theta^{-4}\tilde{p}.
\]
Following the approach used in the discussion of geometric fields, we
will work directly with the frame components $T_{ij}\equiv e_i{}^\mu e_j{}^\nu T_{\mu\nu}$ and
$u_{i}\equiv e_i{}^\mu u_\mu$ with respect to a $g$-orthonormal frame $e_i$. Thus
\[
T_{ij} = (\rho + p )u_i u_j -  p \eta_{ij}.
\]
We observe that $\tilde{g}(\tilde{u},\tilde{u})=1$ implies that
$g(u,u)=1$. Now, using  $u_i=\eta_{ij}u^j$, $u^i =\eta^{ij}u_j$, our signature convention implies
\[
u^0 = u_0, \qquad u^a =-u_a, \qquad a=1,\,2,\,3.
\]

\medskip

A computation using the standard transformation rules for the
covariant derivatives of conformally rescaled metrics yields 
\[
\eta^{ij} \nabla_i T_{jk} = \Theta^{-4}
\tilde{\eta}^{ij}  \tilde{\nabla}_i \tilde{T}_{jk} -
  \Theta^{-5} \tilde{\nabla}_k \Theta\; \tilde{\eta}^{ij} \tilde{T}_{ij}.
\]
Consequently, the (physical) equation for the conservation of
energy-momentum
\[
\tilde{\nabla}^j \tilde{T}_{ij}=0,
\]
implies an analogous equation
\begin{equation}
\nabla^j T_{ij}=0,
\label{UnphysicalEnergyMomentumConservation}
\end{equation}
for the (unphysical) conformally rescaled spacetime only if the
energy-momentum tensor $\tilde{T}_{ij}$ is tracefree ---see \cite{Fri91}.
Notice that $\tilde{T}\equiv \tilde{\eta}^{ij}\tilde{T}_{ij}=0$ if and only if $T\equiv \eta^{ij}
T_{ij}=0$. A quick computation shows that for a perfect fluid the tracefreeness of the
energy-momentum tensor implies $ \rho - 3 p=0$ ---in other words $\gamma=\frac{4}{3}$. Hence  
\[
 p=\tfrac{1}{3} \rho, \qquad \tilde{p}=\tfrac{1}{3}\tilde{\rho}.
\]
This class of perfect fluids is usually referred to as \emph{pure
  radiation}. 

\medskip
\emph{In the present article, our analysis will be restricted to the
  case of tracefree perfect fluids.} The unphysical energy-momentum tensor for
this class of perfect fluids reduces to
\begin{equation}
T_{ij} = \tfrac{4}{3}\rho u_i u_j -  \tfrac{1}{3}  \rho \eta_{ij}.
\label{EMTracelessUnphysical}
\end{equation}

\medskip
As a consequence of the definition of the 4-velocity $u^i$ it follows
that
\begin{subequations}
\begin{eqnarray}
&& \eta_{ij} u^i u^j = u_k u^k = u_0 u^0 + u_a
u^a=1, \label{Norm4Velocity}\\
&& \nabla_k u^0 = -\frac{u_a}{u_0}\nabla_k
u^a, \label{DerivativeNorm4Velocity} \\
&& \nabla_l \nabla_k u^0 =-\frac{u_a}{u_0} \nabla_l \nabla_k u^a -\frac{1}{u_0} \nabla_l u_a \nabla_k u^a -
\frac{u_a u_b}{u_0^3} \nabla_l u^b \nabla_k u^a. \label{DerivativeDerivativeNorm4Velocity}
\end{eqnarray}
\end{subequations}
These identities will be used to rewrite the component $u^0$ and
its derivatives in terms of the spatial components $u^a$ and their
derivatives. This procedure will be central for the construction of a
symmetric hyperbolic system for the matter variables. It is also
noticed that equation \eqref{Norm4Velocity} implies 
\[
u^k \nabla_k \left( u_i u^i \right)=0. 
\]
This expression shows that if $u_i u^i=1$ at some point in a fluid
flow line, then $u_i u^i=1$ in the whole flow line.

\subsection{The energy conservation equation and the equations of motion}

A direct computation shows that the conservation equation
\eqref{UnphysicalEnergyMomentumConservation} implies
\begin{equation}
\label{Fluid_Conservation_Equations}
Z_j \equiv \tfrac{4}{3} \left(u_j u^i \nabla_i \rho + \rho u_j \nabla_i u^i
+ \rho u^i \nabla_i u_j \right) -\tfrac{1}{3} \nabla_j \rho=0.
\end{equation}
This equation can be split into components parallel and orthogonal to
$u^i$:
\begin{subequations}
\begin{eqnarray}
&& u^i Z_i = u^i \nabla_i \rho + \tfrac{4}{3} \rho \nabla_i u^i=0,  \label{ConservationOfEnergy}\\
&& \gamma\tensor{j}{i}{}Z_i = \tfrac{4}{3}\rho u^i \nabla_i u_j+ \tfrac{1}{3} u_j u^i \nabla_i 
\rho -\tfrac{1}{3} \nabla_j \rho =0,  \label{EquationsOfMotion}
\end{eqnarray}
\end{subequations}
where
\[
\gamma_{ij}\equiv \eta_{ij}-u_i u_j . 
\]

These equations are the conformal versions of the \emph{equation of energy conservation}
and the \emph{equations of motion} ---see e.g. \cite{Cho08}. It is
noticed that equations \eqref{ConservationOfEnergy} and
\eqref{EquationsOfMotion} can be combined to give
\begin{equation}
\nabla_j \rho = 4 \rho u^i \nabla_i u_j -\tfrac{4}{3}
\rho u_j \nabla_i u^i.
\label{GradientDensity}
\end{equation}
This equation will be used in the sequel to eliminate the gradient of
the unphysical density from certain expressions.

\subsection{A symmetric hyperbolic system for the fluid fields}
\label{Subsection:FluidHR}

The equations of conservation of energy and motion will be used to
construct a symmetric hyperbolic system of evolution equations for the
unphysical density $\rho$ and the spatial components of the unphysical
velocity $u^a$. The procedure used here follows the presentation given
in \cite{Cho08}. In the sequel, it should be understood that,
consequently with equation \eqref{Norm4Velocity}, 
\[
u_0=u^0=\sqrt{1-u_a u^a}.
\]

\medskip
Substituting identity \eqref{DerivativeNorm4Velocity} into \eqref{ConservationOfEnergy} gives
\begin{equation}
\frac{3}{16\rho^2} \left(u^0 \nabla_0 \rho + u^a \nabla_a \rho \right)+
\frac{1}{4\rho}\left( \nabla_a   u^a - \frac{u_a}{u_0} \nabla_0   u^a  \right)=0,
\label{FluidSHS1}
\end{equation}
where the extra factor $1/4\rho$ has been included to ensure symmetric
hyperbolicity. Similarly, from equation \eqref{EquationsOfMotion} one
deduces
\begin{eqnarray*}
&& \gamma^{0k}Z_k = u^k \nabla_k u^0 + \frac{1}{4\rho} u^0 u^k \nabla_k \rho -
\frac{1}{4\mu} \eta^{0k} \nabla_k \rho=0, \\ 
&& \gamma^{ak}Z_k =u^k \nabla_k u^a + \frac{1}{4\rho} u^a u^k \nabla_k \rho -
\frac{1}{4\rho} \eta^{ak} \nabla_k \rho=0.  
\end{eqnarray*}
In order to obtain suitable evolution equations for the spatial
components of the 4-velocity, we consider the combination
\[
\varsigma^a \equiv\frac{u^a}{u^0} \gamma^{0k}Z_k - \gamma^{ak}Z_k =0,
\]
or equivalently
\begin{equation}
\varsigma^a=\left( u^i \nabla_i u^a + \frac{u^au^i u_c}{u^0u_0} \nabla_i u^c
\right) + \frac{1}{4\rho} \left( \frac{u^a}{u^0}\eta^{0k} \nabla_k
  \rho - \eta^{ak} \nabla_k \rho \right)=0.
\label{FluidSHS2}
\end{equation}

A direct inspection shows that: 

\begin{lemma}
\label{Lemma:SymHypFluid}
Equations \eqref{FluidSHS1} and \eqref{FluidSHS2} constitute a symmetric
hyperbolic system for the fields $\rho$ and $u^a$.
\end{lemma}

One also has that:

\begin{lemma}
\label{Lemma:ConsistencyFluid}
A solution $(\rho,u^a)$ to the evolution equations \eqref{FluidSHS1} and
\eqref{FluidSHS2} implies a solution $(\rho, u_0, u^a)$ to 
equation \eqref{Fluid_Conservation_Equations} with $u_0=\sqrt{1-u_a u^a}$.
\end{lemma}

\medskip
\proof We need to show $Z^j=0$. The definition of $u_0$ implies that
\eqref{DerivativeNorm4Velocity} and
\eqref{DerivativeDerivativeNorm4Velocity} hold.  Now, given a solution
to \eqref{FluidSHS1} and \eqref{FluidSHS2}, the right hand side of
\eqref{FluidSHS1} can be rewritten so as to yield  $u^jZ_j=0
$. Substitution into the left hand side of \eqref{FluidSHS2} gives 
\[
\frac{u^a}{u^0} Z^0 - Z^a =0.
\]
 Contracting with $u_a$ and using \eqref{Norm4Velocity}, as well as
 $u_0 \ge1$ gives first $Z^0=0$ and then $Z^a=0$. Hence a solution to
 \eqref{FluidSHS1} and \eqref{FluidSHS2} satisfies \eqref{Fluid_Conservation_Equations}.

\begin{remark}
{\em Let $u^{AA'}$ denote the spinorial counterpart of the
4-velocity vector $u^\mu$. The spinor $u^{AA'}$ can be split using the
spinor $\tau^{AA'}$ as done in \eqref{u_spacespinors}. This implies
\[
u =\sqrt{2} u_0 =\sqrt{2} u^0, \quad \quad u_{AB} = \sigma^a{}_{AB} u_a,
\]
where $\sigma^a_{AB}$ denote the spatial Infeld-van der Waerden symbols. 
\emph{It follows that \eqref{Fluid_Conservation_Equations} implies a symmetric hyperbolic system for the
spinorial components $u$ and $u_{AB}$.} The explicit form of these
equations will not be required in our subsequent analysis.}
\end{remark}  

\subsection{The Cotton-York tensor of a traceless perfect fluid spacetime}

The matter field quantities feedback into the geometric part of the
conformal field equation through the physical Cotton-York tensor
$\tilde{Y}_{\mu\nu\lambda}$. In what follows, the latter is expressed
in terms of tensors, however the frame and spinor component versions
are easily derived from these equations.

\medskip
For a tracefree energy momentum tensor the physical Schouten tensor is given by 
$\tilde{P}_{ij}=\half \tilde{T}_{ij}$ so that
\[
\tilde{Y}_{ijk}= \tilde{\nabla}_{[i} \tilde{T}_{j]k} .
\]

Rewriting this expression in terms of unphysical
quantities one obtains for $T_{ijk} = \Theta^{-2}
\tilde{Y}_{ijk}$ that 
\begin{equation}
T_{ijk}= \nabla_{[i} T_{j]k} +
 \Upsilon_{[i} T_{j]k} +
 g_{k[i}T_{j]l}\Upsilon^l.
\label{RescaledCYTraceless}
\end{equation}
The last two terms in this expression are polynomial in $\rho$ and the
components $u^i$. The first term, however, contains derivatives of $u^i$ and $\rho$ 
that would enter the principal part of the Cotton-York and Bianchi
equations. The fluid equations cannot be used to eliminate these derivatives.

\medskip
In order to get around this difficulty, we introduce new variables
$\rho_k$ and $u_{ij}$ and corresponding zero quantities $q_k$ and
$y_{ij}$ via 
\begin{equation}
q_k \equiv \rho_k - \nabla_k \rho, \qquad
y_{ij} \equiv u_{ij} -\nabla_i u_j.
\label{Definition:Derivative:rho:and:U}
\end{equation}
Observe that if $q_k=0$ and $y_{ij}=0$, one then has that
$u_{ij}u^j=0$ and $y_{ij}u^j=0$, so that one can write
\[
u_{i0}= -\frac{u^a}{u_0}u_{ia}, \qquad y_{i0}= -\frac{u^a}{u_0}y_{ia}.
\]
Furthermore, from $u_{ij}u^j =0$, it also follows that
\begin{subequations}
\begin{eqnarray}
&& \nabla_k u_i{}^0 = -\frac{u_a}{u_0}\nabla_k u_i{}^a
+\frac{u_k{}^0}{u_0}(y_{i0}-u_{i0})+\frac{u_k{}^a}{u_0}(y_{ia}-u_{ia}), \label{nabla_k
  u_i0} \\
&& u^j\nabla_i u_{kj}= u_{kj}y_i{}^j -u_{kj}u_i{}^j. \label{u nabla u_kj}
\end{eqnarray}
\end{subequations}

\medskip
Finally, if $q_k = 0, \,y_{ij}=0  $, then the first term of $T_{ijk}$ can be written as
\begin{eqnarray*}
&& \nabla_{[i} T_{j]k}
= \tfrac{4}{3}(\nabla_{[i} \rho u_{j]} u_k +  \rho \nabla_{[i}u_{j]} u_k +  \rho u_{[j} \nabla_{i]}u_k)
- \tfrac{1}{3} \nabla_{[i}\rho \eta_{j]k}, \nonumber \\
&& \phantom{\nabla_{[i} T_{j]k}}= \tfrac{4}{3}(\rho_{[i} u_{j]} u_k +  \rho u_{[ij]} u_k +  \rho u_{[j} u_{i]k})
- \tfrac{1}{3} \rho_{[i} \eta_{j]k}.
\end{eqnarray*}

\subsubsection{A symmetric hyperbolic system for $\rho_k$ and $u_k{}^a$}
\label{Subsection:DFluidHR}

The evolution equations for $\rho$ and $u^a$ are derived from equation
\eqref{Fluid_Conservation_Equations}. Taking derivatives of
\eqref{Fluid_Conservation_Equations} and commuting them gives:
\begin{eqnarray}
&& 0=\nabla_k Z_j = 
\tfrac{4}{3} \left(u_j u^i \nabla_i \nabla_k \rho + \rho u_j \nabla_i \nabla_k u^i
+ \rho u^i \nabla_i \nabla_k u_j \right) -\tfrac{1}{3} \nabla_j \nabla_k \rho \nonumber \\
&&\hspace{2cm} +
\tfrac{4}{3} \big(
u_j u^i \Sigma\tensor{k}{l}{i} \nabla_l \rho + \rho u_j r\tensor{ki}{i}{l}u^l - \rho u^i r\tensor{ki}{l}{j} u_l  +
\nabla_k u_j u^i \nabla_i \rho + \nabla_k \rho u_j \nabla_i u^i
\nonumber \\ 
&& \hspace{2cm} + \nabla_k \rho u^i \nabla_i u_j 
u_j \nabla_k u^i \nabla_i \rho + \rho \nabla_k u_j \nabla_i u^i + \rho \nabla_k u^i \nabla_i u_j 
\big) 
-\tfrac{1}{3} \Sigma\tensor{k}{l}{j} \nabla_l  \rho, \nonumber \\
&&\phantom{ 0=\nabla_k Z_j}= 
\tfrac{4}{3} \left(u_j u^i \nabla_i \nabla_k \rho + \rho u_j \nabla_i \nabla_k u^i
+ \rho u^i \nabla_i \nabla_k u_j \right) -\tfrac{1}{3} \nabla_j
\nabla_k \rho + V_{kj} , \label{DZ_j}
\end{eqnarray}
where  all terms with at most one derivative of $\rho$ or $u^k$ have
been gathered in $V_{kj} $. In view of this discussion, in the sequel
we will consider the field equation for $\rho_k$ and $u_{ij}$ given by
the following zero quantity:
\begin{eqnarray}
Z_{kj} \equiv 
\tfrac{4}{3} \left(u_j u^i \nabla_i  \rho_k + \rho u_j \nabla_i  u_k{}^i
+ \rho u^i \nabla_i u_{kj} \right) -\tfrac{1}{3} \nabla_j  \rho_k +
W_{kj} =0, \label{PDE_for_rho_k}
\end{eqnarray}
with
\begin{eqnarray*}
&& W_{kj} \equiv \tfrac{4}{3} \big(
u_j u^i \Sigma\tensor{k}{l}{i} \rho_l + \rho u_j r\tensor{ki}{i}{l}u^l
- \rho u^i r\tensor{ki}{l}{j} u_l   + u_{kj} u^i \rho_i + \rho_k u_j
u_i{}^i + \rho_k u^i u_{ij}  \nonumber \\
&& \hspace{2cm}+u_j u_k{}^i \rho_i + \rho u_{kj}  u_i{}^i + \rho u_k{}^i u_{ij} 
\big) 
-\tfrac{1}{3} \Sigma\tensor{k}{l}{j} \rho_l. 
\end{eqnarray*}
From the equation $Z_{ij}=0$ one derives, in analogy to \eqref{EquationsOfMotion} and \eqref{ConservationOfEnergy}, that
\begin{subequations}
\begin{eqnarray}
&& u^j Z_{kj} \equiv 
u^i \nabla_i  \rho_k + \tfrac{4}{3} \rho \nabla_i  u_k{}^i
+ X_{k} =0, \label{ConservationOfEnergy2}\\
&& \gamma\tensor{j}{l}{}Z_{kl} \equiv  
\tfrac{4}{3} \rho u^i \nabla_i u_{kj} + \tfrac{1}{3} u_j u^i \nabla_i  \rho_k  -\tfrac{1}{3} \nabla_j  \rho_k 
+X_{kj} =0, \label{EquationsOfMotion2}
\end{eqnarray}
\end{subequations}
where \eqref{u nabla u_kj} has been used and
\begin{eqnarray*}
&& X_k \equiv \tfrac{4}{3} \rho u^i (u_{kj}y_i{}^j -u_{kj}u_i{}^j)+ u^j W_{kj} ,\\
&& X_{kj} \equiv -\tfrac{4}{3} \rho u_j u^i (u_{kl}y_i{}^l -u_{kl}u_i{}^l)+ \gamma\tensor{j}{l}{} W_{kl} .
\end{eqnarray*}

\medskip
Finally, we rewrite \eqref{ConservationOfEnergy2} in the form
\begin{equation}
\label{FluidSHS1_k}
\frac{3}{16\rho^2} \left(u^0 \nabla_0 \rho_k + u^a \nabla_a \rho_k \right)+
\frac{1}{4\rho}\left( \nabla_a   u_k{}^a - \frac{u_a}{u_0} \nabla_0   u_k{}^a  \right)+ \hat{X}_k =0
\end{equation}
where 
\[
\hat{X}_k=\frac{3}{16\rho^2} X_k + \frac{u^i}{4\rho}
\left(\frac{u_k{}^0}{u_0}(y_{i0}-u_{i0})+\frac{u_k{}^a}{u_0}(y_{ia}-u_{ia})\right).
\]
Similarly, the combination
\[
 \frac{u^a}{u_0}\gamma^{0l}{}Z_{kl} -\gamma^{al}{}Z_{kl} 
\]
leads to the evolution equation
\begin{equation}
\left( u^i \nabla_i u_k{}^a + \frac{u^a u^i u_c}{u^0u_0} \nabla_i u_k{}^c
\right) + \frac{1}{4\rho} \left( \frac{u^a}{u^0}\eta^{0l} \nabla_l
  \rho_k - \eta^{al} \nabla_l \rho_k \right)+ \hat{X}_k{}^a=0.
\label{FluidSHS2_k}
\end{equation}
where $\hat{X}_k{}^a$ is a combination of
\[
\frac{u^a}{u_0}\gamma^{0l}{}X_{kl} -\gamma^{al}{}X_{kl} 
\]
 and terms from
expression \eqref{nabla_k u_i0}.

\medskip
In analogy to Lemma \ref{Lemma:SymHypFluid} one can readily verify that:

\begin{lemma}
If $y_{ij}=0$, then  equations \eqref{FluidSHS1_k} and \eqref{FluidSHS2_k} constitute a symmetric
hyperbolic system for the fields $\rho_k$ and $u_k{}^a$.
\end{lemma}

A similar argument to the one leading to Lemma
\ref{Lemma:ConsistencyFluid} yields:

\begin{lemma}
\label{Lemma:ConsistencyDFluid}
Let $y_{ij}=0$. A solution $(\rho_k,u_k{}^a)$ to the evolution equations \eqref{FluidSHS1_k} and
\eqref{FluidSHS2_k} implies a solution to equation \eqref{PDE_for_rho_k}.
\end{lemma}

\subsubsection{The subsidiary equations for the fluid variables}

In this section we derive evolution equations for the zero quantities
$q_i$ and $y_{ij}$. These \emph{subsidiary equations} will be of
relevance in the discussion of the propagation of the constraints
---see Section \ref{Section:PropagationConstraints}.

\medskip
Subtracting equation \eqref{DZ_j} from equation \eqref{PDE_for_rho_k} gives
\begin{equation}
 Q_{kj} \equiv  
\tfrac{4}{3} \left(u_j u^i \nabla_i  q_k + \rho u_j \nabla_i  y_k{}^i
+ \rho u^i \nabla_i y_{kj} \right) -\tfrac{1}{3} \nabla_j  q_k +
W_{kj}-V_{kj} =0. \label{PDE_for_q_k}
\end{equation}
Now, using substitutions like 
\[
u_{ij} \rho_k - \nabla_i u_j \nabla_k \rho = u_{ij} q_k + y_{ij}
\rho_k - y_{ij} q_k
\]
one can deduce that all individual terms in sums of $W_{kj}-V_{kj} $
in equation \eqref{PDE_for_q_k} contain at least one zero
quantity. Repeating the discussion for the evolution equations for
$(\rho_k,u_{ij})$ with $(q_k,y_{ij})$ one finds that 
\begin{eqnarray}
&& \frac{3}{16\rho^2} \left(u^0 \nabla_0 q_k + u^a \nabla_a q_k \right)+
\frac{1}{4\rho}\left( \nabla_a   y_k{}^a - \frac{u_a}{u_0} \nabla_0
  y_k{}^a  \right)+ Y_k =0, \label{FluidSHS1_zero_quant}
\\
&& \left( u^i \nabla_i y_k{}^a + \frac{u^a u^i u_c}{u^0u_0} \nabla_i y_k{}^c
\right) + \frac{1}{4\rho} \left( \frac{u^a}{u^0}\eta^{0l} \nabla_l
  q_k - \eta^{al} \nabla_l q_k \right)+ Y_k{}^a=0,
\label{FluidSHS2_zero_quant}
\end{eqnarray}
where all terms in $Y_k $ and $Y_k{}^a $ contain zero quantities. The
evolution equations \eqref{FluidSHS1_zero_quant} and
\eqref{FluidSHS2_zero_quant} constitute a symmetric hyperbolic system
for the independent components of $q_k$ and $y_{ij}$. 

\subsubsection{Final remarks}

As a consequence of the analysis in the previous subsections one has
that the components $T_{ijk}$ of the tensor $T_{\mu\nu\lambda}$ with
respect to the frame $e_i$ are polynomial expressions of the unknowns
$\rho$, $\rho_j$, $u_i$ and $u_{ij}$. If desired, the dependence with
respect to $\rho_k$ can be eliminated using equation
\eqref{GradientDensity}. As long as $\rho\neq 0$, the fields $\rho$, $\rho_j$, $u_i$
and $u_{ij}$ satisfy symmetric hyperbolic equations\footnote{An
  alternative evolution system which is valid even if $\rho=0$ can be
  obtained using the ideas of \cite{Ren92b}.}. Finally, it is noticed
that if the fields $\rho$, $\rho_j$, 
  $u_i$ and $u_{ij}$ are regular at the points where $\Theta=0$, then
  $T_{ijk}$ is also regular ---and consequently, also equations
  \eqref{Cotton-York} and \eqref{Bianchi} are formally regular.

\section{A symmetric hyperbolic reduction of the conformal field equations}
\label{Section:derive_SHsystem}

In the previous section it has been shown how the equations of motion
for the fluid variables and their derivatives lead to a system of
symmetric hyperbolic equations independently of geometric gauge
considerations. The purpose of this section is to briefly discuss a
reduction procedure for the geometric unknowns. Our treatment is
inspired on the one given in \cite{Fri91}, but it also combines ideas
from \cite{Fri86b,LueVal10}. 

\subsection{Gauge freedom}
As mentioned previously, the conformal Einstein field equations
\eqref{XCFEFrame1} and \eqref{XCFEFrame2} are endowed with three
classes of gauge freedom: conformal, coordinate and frame. In what
follows, we briefly discuss a procedure for fixing this freedom.

\subsubsection{Conformal gauge freedom}
As already mentioned, the conformal Einstein field equations
\eqref{XCFEFrame1} and \eqref{XCFEFrame2} admit certain freedom in the
specification of the representative, $g_{\mu\nu}$, of the conformal
class $[\tilde{g}_{\mu\nu}]$ which will be used as the unphysical metric
---see e.g. \cite{Fri03a} and references therein for more details. 

\medskip
Assuming for a moment that one has a solution to the conformal
Einstein field equations with matter, it follows then by contraction
of indices in equation \eqref{transform_Schouten} together with the
tracefreeness  of the energy momentum tensor  that
\[
\nabla_k\nabla^k \Theta = \Theta P_k{}^k.
\]
As discussed in, e.g. \cite{Fri03a}, this equation can always be
solved locally so that the condition
\begin{equation}
P\equiv P_k{}^k=-1, \label{conformal factor}
\end{equation}
holds. This condition fixes the conformal freedom in the
equations \eqref{XCFEFrame1} and \eqref{XCFEFrame2}.

\subsubsection{Coordinate and gauge freedom}

In order to fix the coordinate and frame gauge freedom, we make use of
the notion of \emph{gauge source functions} ---see
\cite{Fri85,Fri91,Fri03a}. The gauge source freedoms will be chosen so
as to render  symmetric hyperbolic evolution equations for the
geometric unknowns.

\medskip
In what follows we encounter equations of the form
\[
2\nabla_{[i}M_{j]\cdots} = N_{[ij]\cdots}
\]
where the dots denote an arbitrary set of indices ---cfr. equations
\eqref{torsionfree}-\eqref{Cotton-York}. The spinorial equivalents of
the above equation are given by
\begin{equation}
\label{DMequ_symm_form}
\nabla_{A(A'}M^A{}_{B')\cdots} = N_{A(A'}{}^A{}_{B')\cdots},
\end{equation}
and its complex conjugate. Now, an equation of the form
\begin{equation}
\label{DMequ}
\nabla_{AA'}M^A{}_{B'\cdots} = N_{AA'}{}^A{}_{B'\cdots}
\end{equation}
is well known to imply a symmetric hyperbolic evolution system for the
independent components of $M_{AA'}$ ---see e.g. \cite{Fri85}.  Note,
however, that equation \eqref{DMequ_symm_form} contains no information about
the skew term
\begin{equation}
\nabla_{A[A'}M^A{}_{B']\cdots} = \tfrac{1}{2}\epsilon_{A'B'}
\nabla^{CC'}M_{CC'\cdots} =F_{\cdots},
\label{SkewTerm}
\end{equation}
which can be specified arbitrarily. Thus, by adding \eqref{SkewTerm}
with a convenient choice of a gauge source function, $F_{\cdots}$, to
\eqref{DMequ_symm_form} one obtains an equation of the form of
\eqref{DMequ}, from where a symmetric hyperbolic system can be
extracted ---see e.g. \cite{Fri85,Fri03a}.

\medskip
The previous discussion will be implemented in the field equations
\eqref{torsionfree}, \eqref{Curvature} and \eqref{Cotton-York}. These
equations provide differential conditions for the fields $e^\mu_i$,
$P_{ij}$ and $\Gamma_i{}^j{}_k$. Let $e^\mu_{AA'}$, $P_{AA'BB'}$ and
$\Gamma_{AA'}{}^{BB'}{}_{CC'}$ denote the spinorial counterparts of
these fields. As a consequence of the metricity of the connection,
instead of working with $\Gamma_{AA'}{}^{BB'}{}_{CC'}$, we will
consider a spinorial field $\Gamma_{AA'BC}$ such that
\[
\Gamma_{AA'}{}^{BB'}{}_{CC'} = \Gamma_{AA'}{}^B{}_C
\epsilon_{C'}{}^{B'} + \bar{\Gamma}_{A'A}{}^{B'}{}_{C'}\epsilon_C{}^{B}.
\]
For convenience, define the gauge source functions
\begin{subequations}
\begin{eqnarray}
&& F^\mu\equiv \nabla^{AA'} e^\mu_{AA'},  \label{Definition:GaugeSource1}\\
&& F_{(BC)} \equiv \nabla^{AA'} \Gamma_{AA'BC},  \\
&& F_{BB'}  \equiv \nabla^{AA'} P_{AA'BB'} = \nabla_{BB'}P, \label{Definition:GaugeSource3}
\end{eqnarray}
\end{subequations}
where the second equality in the definition of $F_{BB'}$ follows from
the  twice contracted Bianchi identity for the unphysical connection
$\nabla$.  Motivated by their value in the reference solution (the
conformal FLRW solution) the gauge source functions will be fixed by the
conditions
\begin{equation}
F^\mu=0, \qquad F_{(BC)}=0, \qquad F_{BB'}=0.
\label{Choice:gaugeSorceFunctions}
\end{equation}
Notice, in particular, that the last condition is consistent with the
conformal gauge condition \eqref{conformal factor}. As discussed in
\cite{Fri85} ---see also \cite{Fri91}--- a particular choice of the
coordinate and frame gauge functions $F^\mu$ and $F_{(BC)}$ fixes the
coordinates and frame\footnote{The gauge source functions $F^\mu$ and
  $F_{(BC)}$ imply, respectively, wave equations for the coordinates
  and a semi-linear equation for the frame components. These equations
  can be solved locally.}.

\medskip
The spacetimes to be considered in the present analysis have the
topology of $\Real \times \Sphere^3$. Given an initial manifold
$\mathcal{S}$ for the spacetime, then there is a diffeomorphism $\Phi:
\mathcal{S} \rightarrow \Sphere^3$ which allows to pull-back
coordinates from $\Sphere^3$ to $\mathcal{S}$. These coordinates on
the initial manifold $\mathcal{S}$ will be used as the initial value
of the spatial part of the spacetime coordinates. The time coordinate
will be set initially to zero. The initial value of the frame $e_i$ is
set by choosing on $\mathcal{S}$ some arbitrary orthonormal 
spatial frame $e_a$  (with respect to the
3-metric of $\mathcal{S}$). The $e_0$ vector is set to coincide
initially with the (spacetime) normal to $\mathcal{S}$.

\subsection{The evolution equations}
\label{Subsection:SH_system}

The hyperbolic reduction of the matter variables has already been
discussed in Sections \ref{Subsection:FluidHR} and
\ref{Subsection:DFluidHR}. In what concerns the evolution equations
for the geometric variables, we follow the procedure indicated in
\cite{Fri91}. This consists of a rewriting the spinorial version of the
conformal field equations \eqref{XCFEFrame1}-\eqref{XCFEFrame2} in
terms of space spinors so that the resulting equations contain only
unprimed indices. In order to encompass the full information of the
field equations, one has to include into the set of equations their
Hermitian conjugates. If the fields and equations are then decomposed into
their irreducible parts, then the equations split in a natural way
into symmetric hyperbolic evolution and constraint equations. This
procedure is straightforward, but involves lengthy computations, most of which
can now be implemented in a computer algebra system like the suite
{\tt xAct} for {\tt Mathematica}\footnote{See {\tt www.xAct.es.}}. 

\medskip
The required evolution equations have already been deduced in
\cite{Fri91}. Their detailed form will not be required here. Instead
we present a summary of their key structural properties. In what follows let
\begin{eqnarray*}
&& {\bm \upsilon} \equiv \left( \Theta, d_{AB}, s, e^\er_{AB},
  \Gamma_{ABCD}, P_{ABCD}\right),\\
&& {\bm \phi} \equiv \phi_{ABCD}, \\
&& {\bm \varrho} \equiv\left( \rho, u_{(AB)} \right), \\
&& {\bm \psi} \equiv\left( \rho_{AB}, u_{AB(CD)}  \right),
\end{eqnarray*}
where only the independent irreducible components of the spinors are taken into
account. In terms of these objects, the evolution equations have the
form
\begin{subequations}
\begin{eqnarray}
&& \partial_0 {\bm \upsilon} + {\bm A}_{[\bm \upsilon]}^\er ({\bm
  \upsilon})c_\er {\bm \upsilon} = 
{\bm B}_{[\bm \upsilon]} ({\bm \upsilon}){\bm \upsilon} +
{\bm M}_{[\bm \upsilon]} ({\bm \upsilon},{\bm \phi},{\bm \varrho},{\bm \psi}), \label{Reduced1}\\
&& \left( \sqrt{2}{\bm E} +{\bm A}^{\underline{0}}_{[\bm\phi]}( {\bm \upsilon})\right) \partial_0{\bm\phi}
+ {\bm A}^\er_{[\bm\phi]}( {\bm \upsilon})c_\er {\bm
  \phi}={\bm B}_{[\bm\phi]} ({\bm\upsilon}){\bm\phi} +
{\bm M}_{[\bm\phi]} ({\bm \upsilon},{\bm\psi},{\bm\varrho}), \label{Reduced2}\\
&&{\bm A}^{\underline{0}}_{[\bm\varrho]}({\bm\upsilon},{\bm\varrho}) \partial_0 {\bm\varrho} +
{\bm A}^\er_{[\bm\varrho]}({\bm\upsilon},{\bm\varrho})c_\er{\bm\varrho}
={\bm B}_{[\bm\varrho]}({\bm\upsilon}){\bm\varrho}\label{Reduced3}, \\
&& {\bm A}^{\underline{0}}_{[\bm\psi]}({\bm\upsilon},{\bm\varrho}) \partial_0 {\bm\psi} +
{\bm A}^\er_{[\bm\psi]}({\bm\upsilon},{\bm\varrho})c_\er{\bm\psi}
={\bm B}_{[\bm\psi]}({\bm\upsilon}){\bm\psi} + {\bm M}_{[\bm\psi]}({\bm\upsilon},{\bm\phi},{\bm\varrho}) \label{Reduced4}.
\end{eqnarray}
\end{subequations}
In equations \eqref{Reduced1}-\eqref{Reduced4}, $\mathbf{E}$ denotes
the $5\times 5$ unit matrix, while ${\bm A}^{\es}_{[\bm\phi]}$,
${\bm A}^\es_{[\bm\varrho]}$,
${\bm A}^\es_{[\bm\psi]}$ denote smooth
symmetric matrix-valued functions of their respective arguments. In
particular,  ${\bm A}^{\underline{0}}_{[\bm\phi]} (
{\bm 0})={\bm 0}$ and ${\bm A}^{\underline{0}}_{[\bm\varrho]}$,
${\bm A}^{\underline{0}}_{[\bm\psi]}$ are positive
definite if $\rho>0$. In addition, ${\bm B}_{[\bm \upsilon]}$,
${\bm B}_{[\bm\phi]}$,
${\bm B}_{[\bm\varrho]}$ and ${\bm B}_{[\bm\psi]}$ are smooth
matrix-valued functions of $\bm \upsilon$. Finally, ${\bm M}_{[\bm
  \upsilon]}$, ${\bm M}_{[\bm\phi]}$ and ${\bm M}_{[\bm\psi]}$
are non-linear vector-valued functions of their respective
arguments. These functions are 
smooth if $u^0\neq 0$. 

\medskip
For convenience of the discussion, in the sequel, we define
\[
{\bm w}\equiv
(\mbox{Re}({\bm \upsilon}), \mbox{Im}({\bm\upsilon}), 
\mbox{Re}({\bm \phi}), \mbox{Im}({\bm\phi}),
\mbox{Re}({\bm \varrho}), \mbox{Im}({\bm \varrho}), 
\mbox{Re}({\bm \psi}), \mbox{Im}({\bm \psi})).
\]

\begin{remark}
{\em The deduction of the evolution equations
\eqref{Reduced1}-\eqref{Reduced4} assumes the choice of gauge source functions
given in \eqref{Choice:gaugeSorceFunctions}.} 
\end{remark}

\subsection{Propagation of the constraints}
\label{Section:PropagationConstraints}

An analysis of the so-called \emph{subsidiary equations} describing
the propagation of the zero quantities
\begin{equation}
\Sigma_i{}^l{}_j, \quad \Xi^k{}_{lij}, \quad \Delta_{lij}, \quad
\Lambda_{lij}, \quad \delta_k, \quad \delta_{ij}, \quad \zeta_k, \quad \zeta,
\label{List:ZeroQuantities}
\end{equation}
in terms of which the geometric part of the conformal field equations
---see equations \eqref{XCFEFrame1}-\eqref{XCFEFrame2}---
is expressed has been given in \cite{Fri91}. This lengthy analysis is
succinctly summarised in the following lemma:

\begin{lemma}
\label{Lemma:SubsidiaryEquations}
If the unphysical energy-momentum tensor, $T_{ij}$, satisfies
\[
\nabla^i T_{ij}=0
\]
 and the expressions 
\begin{equation}
\nabla^i T_{ijk}, \qquad \left( \tfrac{1}{2} \Theta^3 T_i{}^m d_{mjkl} + \nabla_i \Theta
  T_{klj} + \Theta \nabla_i T_{klj} \right)\epsilon^{ikl}{}_n - d^m
\epsilon^{kl}{}_{mj} T_{kln},
\label{LooseTerms}
\end{equation}
can be rewritten in terms of matter zero quantities, then the
geometric zero quantities in \eqref{List:ZeroQuantities}  satisfy a
subsidiary system which is symmetric hyperbolic and homogeneous in the
zero quantities.
\end{lemma}

\medskip
A lengthy computation assuming the form for the energy-momentum tensor
given by equation \eqref{EMTracelessUnphysical} and taking into
account the expression \eqref{RescaledCYTraceless} for the
rescaled Cotton-York tensor, shows that the expressions
\eqref{LooseTerms} in Lemma \ref{Lemma:SubsidiaryEquations} can be
rewritten as a homogeneous expressions of the matter zero quantities
$q_k$, $y_{ij}$, $Z_{kj}$ and $Q_{kj}$ defined by equations
\eqref{Definition:Derivative:rho:and:U}, \eqref{PDE_for_rho_k} and
\eqref{PDE_for_q_k}, 
respectively. The analogue of Lemma \ref{Lemma:SubsidiaryEquations} for the matter
zero quantities is given by Lemmas \ref{Lemma:ConsistencyFluid} and
\ref{Lemma:ConsistencyDFluid}. 

\medskip
The purpose of the analysis of the propagation of the constraints is
to establish the following \emph{reduction theorem} which follows directly
from the symmetric hyperbolicity of the subsidiary systems and their
homogeneity with respect to the zero quantities.

\begin{theorem}
\label{Reduction:Theorem}
A smooth solution $\bm w$ of the propagation equations
\eqref{Reduced1}-\eqref{Reduced2} which satisfies the constraint
equations on a spacelike hypersurface $\mathcal{S}$ defines in the
domain of dependence of $\mathcal{S}$ a solution to the conformal
Einstein field equations with matter model given by a traceless
perfect fluid.
\end{theorem}

\section{The traceless perfect fluid FLRW cosmology as a solution to the conformal
  Einstein field equations}
\label{FLRW:CFE}

The purpose of the present section is to cast the traceless perfect
fluid FLRW cosmology with $\lambda<0$ in a form in which its character
as a solution to the conformal Einstein field equations with matter
becomes manifest. 

\subsection{The FLRW cosmology on the Einstein cylinder}
\label{Subsection:FLRWEinsteinCylinder}

One of the most important properties characterising FLRW cosmologies
is their conformal flatness. This shows that as in the case of the
Minkowski, de Sitter and anti-de Sitter spacetimes, these solutions
admit a conformal representation in which the unphysical spacetime
$(\mathcal{M},g_{\mu\nu})$ is given by the so-called \emph{Einstein
cylinder (or Einstein cosmos)}.

\medskip
The Einstein cosmos is given by the manifold
$\mathcal{M}_{\mathscr{E}}=\Real \times \Sphere^3$ with a metric given
by the line element
\begin{equation}
\label{EC_metric}
g_{\mathscr{E}}= \mbox{d}\tau ^2 - \mbox{d}\sigma^2
\end{equation}
where, again, $d\sigma^2$ is the standard line element of
$\Sphere^3$. The manifold $\Sphere^3$ will be coordinatised in the way
indicated in Subsection \ref{Subsection:Coordinates3Sphere}. A
$g$-orthonormal frame $\mathring{e}_k$ can be defined on
$\mathcal{M}_{\mathscr{E}}$ by completing the frame $\{c_1, \, c_2, \, c_3\}$
discussed also in Subsection \ref{Subsection:Coordinates3Sphere} with
the vector $\mathring{e}_0=\partial_\tau$. Setting, for convenience,
$c_0\equiv \partial_\tau$, one can write $\mathring{e}_k =
\mathring{e}_k{}^\es c_\es$ with $\mathring{e}_k{}^\es
=\delta_k{}^\es$ ---the components of $\mathring{e}_k$ with respect to the basis
$c_\es$.

\medskip
In order to relate the FLRW line element \eqref{FRW:AltLineElement}
with that of the Einstein cosmos, equation \eqref{EC_metric}, one
introduces the change of coordinate
\[
\tau = \int_{t_0}^t \frac{\mbox{d}s}{a(s)}.
\]
This naturally leads to the following choice of conformal factor:
\begin{equation}
\mathring{\Theta}(\tau)\equiv1/a(\tau) \equiv 1/a(t(\tau)),
\label{ConformalFactor:FLRW}
\end{equation}
so that $g_{\mathscr{E}}=\mathring{\Theta}^2 \tilde{g}_{\mathscr{F}}$, 
where $\tilde{g}_{\mathscr{F}}$ is given by equation
\eqref{FRW:AltLineElement}. For the class of FLRW cosmologies covered
by Proposition \ref{Proposition:ScaleFactor} one has that
$\mathring{\Theta}\rightarrow 0$ as $t\rightarrow \infty$. Moreover,
there exists a (finite) positive constant $\tau_\infty$ such that
$\mathring{\Theta}(\tau_\infty)=0$. Notice also that $\tau=0$ for
$t=t_0$.

\medskip
A direct computation using the line element \eqref{EC_metric}, the
frame $\mathring{e}_k$ and the conformal factor
\eqref{ConformalFactor:FLRW} gives the following expressions for the
unknowns of the conformal field equations:
\begin{subequations}
\begin{eqnarray}
&& \mathring{e}_k{}^\es=\delta_k{}^\es, \qquad \mathring{\Gamma}\tensor{i}{j}{k} = \epsilon_{0ilk}\eta^{jl}, \quad \mathring{P}_{ij} = \delta_i{}^0 \delta_j{}^0 - \tfrac{1}{2} \eta_{ij}, \quad \mathring{d}_{ijkl}=0, \label{ReferenceSolution1}\\
&& \mathring{\Theta} = a^{-1}, \quad \mathring{d}_k=-a^{-2}a'\delta_k{}^0, \quad  \mathring{s}= \tfrac{1}{2} a^{-3} a^{\prime 2}-\tfrac{1}{4} a^{-2} a^{\prime\prime} -\tfrac{1}{4}a^{-1}, \label{ReferenceSolution2} \\
&& \mathring{\rho} = \tilde{\rho}_0 a_0^4, \quad \mathring{u}_i = \delta_i{}^0, \quad \mathring{\rho}_k=0, \quad \mathring{u}_{ij} =0. \label{ReferenceSolution3}
\end{eqnarray}
\end{subequations}
where ${}'$ denotes differentiation with respect to $\tau$, and $a_0$ and $\tilde{\rho}_0$
are the values of the scale factor and the physical pressure at the initial time $\tau=0$. 
Notice that the unphysical density for this model is constant. Spinorial versions of the
above expressions can be readily obtained by contraction with the
constant spacetime Infeld-van der Waerden symbols $\sigma^i{}_{AA'}$,
or their space spinor version $\sigma^a{}_{AB}$. The explicit
expressions will not be required in our subsequent analysis. Following
the notation of Lemma \ref{Lemma:SHReduction}, we collect the
independent spinorial components of the fields in
\eqref{ReferenceSolution1}-\eqref{ReferenceSolution3} in a vectorial
unknown which we denote by $\mathring{\bm w}$. 

\medskip
A direct computation using the expressions
\eqref{ReferenceSolution1}-\eqref{ReferenceSolution3} shows that, for
this solution, the
gauge source functions $F^\mu$, $F_{(AB)}$ and $F_{AA'}$ as defined by
\eqref{Definition:GaugeSource1}-\eqref{Definition:GaugeSource3} are
given by
\[
F^\mu=0, \quad F_{(AB)}=0, \quad F_{AA'}=0.
\]
This computation justifies the choice of gauge source functions made
in \eqref{Choice:gaugeSorceFunctions}.

\medskip
Recalling that $\mbox{d}/\mbox{d}\tau = a \mbox{d}/\mbox{d}t$,
and using the limits given in Proposition
\ref{Proposition:ScaleFactor}, it follows that
\begin{equation}
\mathring{d}_k \rightarrow -\sqrt{-\tfrac{1}{3}\lambda}, \quad \mathring{s}\rightarrow 0 \quad \mbox{ as } \tau\rightarrow \tau_\infty.  
\label{Limits}
\end{equation}
Accordingly, the expressions given in \eqref{ReferenceSolution1},
\eqref{ReferenceSolution2} and \eqref{ReferenceSolution3} define a
smooth solution to the conformal Einstein field equations
\eqref{XCFEFrame1}-\eqref{XCFEFrame2} for $\tau\in[0,\tau_\infty]$. In
fact, this solution extends, at least locally, beyond
$\tau=\tau_\infty$. This can be easily seen to be the case by using
the values of the solution
\eqref{ReferenceSolution1}-\eqref{ReferenceSolution3} as the initial
value for a Cauchy problem on the slice $\tau=\tau_\infty$. 
The symmetric hyperbolicity of the evolution equations implies that 
the solution to this initial value problem exists for
$\tau\in[\tau_\infty,\tau_\infty+\delta)$ for some $\delta>0$.
From the expression for
$\mathring{d}_k$ in \eqref{ReferenceSolution2} and Proposition
\ref{Proposition:ScaleFactor} it follows that $\mathring{d}_0<0$ at
$\tau_\infty$. Thus, by continuity, $\delta$ can be chosen such that 
$\mathring{\Theta}<0$ on $(\tau_\infty,\tau_\infty+\delta)$. 
In summary we have:

\begin{lemma}
There exists $\delta>0$ such that the expressions
\eqref{ReferenceSolution1}-\eqref{ReferenceSolution3} give rise to a
solution to the evolution equations implied by the conformal Einstein
field equations \eqref{XCFEFrame1}-\eqref{XCFEFrame2} on
$[0,\tau_\infty+\delta)$. Furthermore $\mathring{\Theta}<0$ in $(\tau_\infty,\tau_\infty+\delta)$.
\end{lemma}

\subsubsection*{Initial data for a FLRW cosmology}

The expressions in
\eqref{ReferenceSolution1}-\eqref{ReferenceSolution3} naturally induce
an initial data set for the conformal Einstein field equations which
we denote by $\mathring{\bm w}_0$. 
Notice that there is no need for performing a
pull-back in this construction as the fields in
\eqref{ReferenceSolution1}-\eqref{ReferenceSolution3} are all scalars.

\subsection{Structure of the conformal boundary}

The structure of the conformal boundary for the solution to the
conformal Einstein field equations described by
\eqref{ReferenceSolution1}-\eqref{ReferenceSolution3} follows directly
by inspection.

\medskip
By construction at $\tau=\tau_\infty$ one has that
$\mathring{\Theta}=0$. From the limits \eqref{Limits} ---see also
equation \eqref{Equation:lambda}--- one has that:
\[
\mathring{d}_k \mathring{d}^k= \nabla_k \mathring{\Theta} \nabla^k \mathring{\Theta} = -\tfrac{1}{3}\lambda >0, \quad \mbox{ at } \quad \tau=\tau_\infty,
\] 
so that the future conformal boundary $\mathscr{I}^+ \equiv\{ p\in
\mathcal{M}_{\mathscr{E}} \;| \; \mathring{\Theta}=0\}$ is spacelike.

\section{Existence and stability results}
\label{Section:Existence}

The purpose of the present section is to provide our main
results. These concern the global existence of solutions to conformal
Einstein field equations with matter source given by a perfect fluid
in the case $\lambda<0$, $\gamma=\tfrac{4}{3}$ which can be regarded as
non-linear perturbations of the reference FLRW solution described by
Proposition \ref{Proposition:ScaleFactor}. We also provide results
concerning the structure of the conformal boundary for these
solutions. Altogether these results show the non-linear stability
towards the future of the reference FLRW cosmological model.

\subsection{An Ansatz for the solution}

We will consider solutions to the evolution equations
\eqref{Reduced1}-\eqref{Reduced4} of the form ${\bm w} = \mathring{\bm
w} + \breve{\bm w}$, where $\mathring{\bm w}$ as defined in Section
\ref{Subsection:FLRWEinsteinCylinder}, and $\breve{\bm w}$ describes a
non-linear perturbation from the reference solution $\mathring{\bm
w}$. The fields in $\mathring{\bm w}$ are interpreted as the pull-back
of the original fields on $\mathcal{M}_{\mathscr{E}}$ under a cylinder map. In
what follows let $\bm w_0= \mathring{\bm w}_0 + \breve{\bm
w}_0$ be an initial data set for the system of evolution equations
\eqref{Reduced1}-\eqref{Reduced4} prescribed on the initial manifold
$\mathcal{S}$ ---as discussed in Subsection
\ref{Subsection:InitialData}. It will be assumed that ${\bm w}_0$
satisfies the conformal constraint equations. The vector $\mathring{\bm
w}_0$ is to be interpreted as the pull-back of the smooth map relating
$\Sphere^3$ and the initial manifold $\mathcal{S}$.

\medskip
A direct inspection gives rise to the following result:
\begin{lemma}
\label{Lemma:SHReduction}
For $\breve {\bm w}$ sufficiently close to $\bm 0$ and as long as
$\rho>0$ and $u^0\neq 0$, the equations \eqref{Reduced1}-\eqref{Reduced4} imply a symmetric
hyperbolic evolution system
\begin{equation}
\label{Prototype:SHS}
{\bm A}^0( \mathring{\bm w} +
\breve{\bm w}) \cdot \partial_\tau \breve{\bm w}+ \sum^3_{\er=1} {\bm A}^\er( \mathring{\bm w} +
\breve{\bm w})\cdot c_\er(\breve{\bm w}) + {\bm
  B}(\tau,{\bm x},\mathring{\bm w},c_\es\mathring{\bm w},\breve{\bm w})\cdot \breve{\bm w}=0,
\end{equation}
 for the independent components of
${\bm w}$. The
matrix valued functions ${\bm A}^\es$, ${\bm B}$ are smooth functions
of their arguments. Furthermore, the entries of the matrix ${\bm
  A}^0(\mathring{\bm w})$ are bounded from below by
$1/\sqrt{2}$. Finally, $\breve{\bm w}={\bm 0}$ is a solution of
equation \eqref{Prototype:SHS}.
\end{lemma}

\subsection{Constructing initial data for the conformal evolution
  equations}
\label{Subsection:InitialData}

In the sequel, it will be assumed that one has a solution $(\mathcal{S},
\tilde{h}_{\alpha\beta}, \tilde{K}_{\alpha\beta}, \tilde{\rho},
\tilde{u}^\alpha)$ to the (physical) $\lambda<0$ Einstein-perfect fluid constraint equations 
\begin{subequations}
\begin{eqnarray}
&& \tilde{r} + \tilde{K}^2 - \tilde{K}_{\alpha\beta} \tilde{K}^{\alpha\beta}=
2(\lambda - \tilde{\mu}) , \label{HamiltonianConstraint} \\
&& \tilde{D}^\alpha \tilde{K}_{\alpha\beta} - \tilde{D}_\beta
\tilde{K} =\tilde{j}_\beta,  \label{MomentumConstraint}
\end{eqnarray}
\end{subequations}
with $\mathcal{S}$ having the topology of $\Sphere^3$ and the perfect
fluid satisfying a barotropic equation of state with $\gamma=\tfrac{4}{3}$.  In
equations \eqref{HamiltonianConstraint}-\eqref{MomentumConstraint}
$\tilde{D}_\beta$ and $\tilde{r}$ denote the
Levi-Civita covariant derivative and the Ricci scalar of the intrinsic
3-metric $\tilde{h}_{\alpha\beta}$ of $\mathcal{S}$.
$\tilde{K}_{\alpha\beta}$ is a symmetric 3-dimensional tensor
corresponding to the extrinsic curvature of $\mathcal{S}$ with respect
to the $\tilde{g}$-unit normal $\tilde{n}_\mu$. Furthermore,
$\tilde{\mu}\equiv \tilde{n}^\mu \tilde{n}^\nu \tilde{T}_{\mu\nu}$,
while $\tilde{j}_\beta$ corresponds to the pull-back to $\mathcal{S}$ of 
$\tilde{j}_\lambda \equiv \tilde{n}^\mu \tilde{h}_\lambda{}^\nu
\tilde{T}_{\mu\nu}$. For a perfect fluid with a tracefree energy-momentum tensor, a
direct computation gives that
\[
\tilde{\mu} = \tfrac{1}{3} \tilde{\rho} (4 \tilde{u}_\parallel -1),
\quad 
\quad \tilde{j}_\beta = \tfrac{4}{3} \tilde{\rho} \tilde{u}_\parallel
\tilde{u}_\beta, 
\]
where $\tilde{u}_\parallel \equiv \tilde{u}^\mu \tilde{n}_\mu$ and $\tilde{u}_\beta$
corresponds to the pull-back of $\tilde{h}_\mu{}^\nu \tilde{u}_\nu$ to
$\mathcal{S}$. In particular, if on $\mathcal{S}$ one has that
$\tilde{n}_\mu$ and $\tilde{u}^\nu$ are aligned ---as in the case of
the FLRW cosmologies--- then
$\tilde{\mu}=\tilde{\rho}$ and $\tilde{j}_\beta=0$. In general,
however, we will consider perfect fluid configurations for which
$\tilde{n}_\mu$ and $\tilde{u}^\nu$ are not aligned.

\medskip
Using a generalisation of the procedure for vacuum spacetimes
described in, say, \cite{Fri83} one can construct a solution to the
conformal constraint equations implied on $\mathcal{S}$ by equations
\eqref{XCFEFrame1}-\eqref{XCFEFrame2}. Following the notation
introduced in Section \ref{Subsection:SH_system} we denote the
independent components of such a solution by ${\bm w}_0$. 

\subsection{The main result}

In what follows, given $m\in \mathbb{N}$, let $||\cdot||_m$ denote the Sobolev-like norm on the
space $C^\infty(\Sphere^3,\Real^N)$ of smooth $\Real^N$ valued
functions on $\Sphere^3$ for some non-negative integer $N$ ---see
e.g. \cite{Fri86a,LueVal09a} for precise definitions. Furthermore, let
$H^m(\Sphere^3,\Real^N)$ be the Hilbert space obtained as the
completion of the space $C^\infty(\Sphere^3,\Real^N)$ in the norm
$||\cdot||_m$. Using the cylinder map between $\mathcal{M}_{\mathscr{E}}$ and
$\mathcal{M}$, one can apply the norm $||\cdot||_m$ to evaluate the
norm of functions on the unphysical initial hypersurface
$\mathcal{S}$. Furthermore, the vector $\breve{\bm w}=\breve{\bm
 w}(\tau,{\bm x})$ can be regarded as a function of $\tau$ which
takes values in $H^m(\Sphere^3,\Real^N)$. 

\medskip
Our main result is the following theorem:

\begin{theorem}
\label{MainTheorem}
Suppose $m\geq 4$. Let $\mathcal{S}$ denote a 3-dimensional manifold
diffeomorphic to $\Sphere^3$, and let ${\bm
w}_0=\mathring{\bm w}_0+\breve{\bm w}_0$ be initial data for the conformal
evolution equations \eqref{Reduced1}-\eqref{Reduced4} constructed from
some physical initial data set, $(\mathcal{S},
\tilde{h}_{\alpha\beta}, \tilde{K}_{\alpha\beta}, \tilde{\rho},
\tilde{u}^\alpha)$, for the Einstein field equations with $\lambda<0$
and matter source given by a traceless perfect fluid
$(\gamma=4/3)$. There exists $\varepsilon>0$ such that if
$||\breve{\bm w}_0||_m<\varepsilon$ then the initial data set ${\bm
w}_0$ determines a unique solution, $\bm w$, to the evolution
equations \eqref{Reduced1}-\eqref{Reduced4} which exists on
$[0,\tau_*]$ with $\tau_*>\tau_\infty$. The solution ${\bm w}$ is of
class $C^{m-2}([0,\tau_*]\times \Sphere^3)$ is such that:

\begin{itemize}
\item[(i)] it determines, in turn, a $C^{m-2}$ solution to the
$\lambda<0$ conformal Einstein field equations, equations
\eqref{XCFEFrame1}-\eqref{XCFEFrame2} and
\eqref{Fluid_Conservation_Equations}-\eqref{PDE_for_rho_k}, with
matter given by a traceless fluid on $[0,\tau_*]\times \Sphere^3$;

\item[(ii)] there exists a function $\tau_+=\tau_+(\bm x)$, ${\bm x} \in \Sphere^3$, such that $0<\tau_+({\bm x}) < \tau_*$ and
\begin{eqnarray*}
&& \Theta>0, \quad \mbox{ on } \quad \tilde{\mathcal{M}}\equiv\{(\tau,{\bm x})\in \Real \times \Sphere^3 \; | \; 0\leq \tau <\tau_+({\bm x})   \}, \\
&& \Theta=0,\quad d_k d^k =-\tfrac{1}{3}\lambda <0 \quad \mbox{ on } \quad  \mathscr{I}^+\equiv \{ (\tau_+({\bm x}),{\bm x})\in \Real \times \Sphere^3 \; |\; {\bm x} \in \Sphere^3 \}.
\end{eqnarray*}

\item[(iii)] one obtains a $C^{m-2}$ solution
$(\tilde{\mathcal{M}},\tilde{g}_{\mu\nu},\tilde{\rho},\tilde{u}^\mu)$, to the $\lambda<0$
Einstein-perfect fluid field equations with $\gamma=4/3$ which is
future geodesically complete for which $\mathscr{I}^+$ as defined
above represents conformal future infinity;

\item[(iv)] given a sequence of initial data ${\bm w}^{(n)}_0$ such
  that $||\breve{\bm w}_0^{(n)}||_m<\varepsilon$ and $||\breve{\bm
      w}^{(n)}_0||_m\rightarrow 0$ as $n\rightarrow \infty$, then for
    the corresponding solutions $\breve{\bm w}^{(n)}$ (with minimum
    existence time $\tau_*$) one has
    $||\breve{\bm w}^{(n)}||_m\rightarrow 0$ uniformly in
    $\tau\in [0,\tau_*]$. 

\end{itemize}
\end{theorem}

\begin{remark}
{\em The above theorem, and in particular part (iv), amounts to a non-linear
stability result for the $\lambda<0$, $\gamma=\tfrac{4}{3}$ FLRW cosmological
models, in the sense that sufficiently small perturbations of data for
the FLRW solution give rise to (future) global solutions to the
Einstein field equations with the same asymptotic structure as the
reference solution.}
\end{remark}

\begin{remark}
{\em Note that no consideration of the vorticity of the radiation
fluid was required for the derivation. The vorticity of the fluid can
be calculated from the components of $u_{ij}$.}
\end{remark}

\begin{proof}
Existence, uniqueness and the smoothness of the solutions to equations
\eqref{Reduced1}-\eqref{Reduced4} follow from the properties of the
equation \eqref{Prototype:SHS} provided in Lemma
\ref{Lemma:SHReduction} and an extension of the general existence and
stability Theorem by Kato \cite{Kat75} provided in \cite{Fri86a}
---see also \cite{LueVal09a}. In particular, if $\varepsilon$ is
sufficiently small, one obtains a common existence time
$\tau_*>\tau_\infty$ for all initial data with $||\breve{\bm
w}_0||_m<\varepsilon$. Part (iv) follows from the same result. 

\medskip
Now, the \emph{Reduction Theorem}, Theorem \ref{Reduction:Theorem}
ensures that if the conformal constraint equations are satisfied on
$\mathcal{S}$, one obtains a $C^{m-2}$ solution on $[0,\tau_*]\times
\Sphere^3$ to the $\lambda<0$ conformal Einstein perfect fluid
equations with $\gamma=\tfrac{4}{3}$ ---this shows part (i).

\medskip
In order to show part (ii), one observes that because $\mathring{\Theta}<0$
in $(\tau_\infty,\tau_\infty+\delta)$, then if $\varepsilon$ is
sufficiently small one has that $\Theta<0$ at, say, $\tau=\tau_\infty
+\delta/2$. As $\Theta>0$ at $\tau=0$, then there is a $\tau_+$ for
which $\Theta=0$. By reducing, if necessary $\varepsilon$ one has that
such $\tau$ is unique, and hence, the function $\tau_+({\bm x})$ is well defined.
If follows from \eqref{1-form} and \eqref{Equation:lambda} 
that $\Theta=0$ implies $d_k d^k=\nabla_k \Theta \nabla^k \Theta = -\frac{1}{3}\lambda >0$. 
Hence $\Theta=0$, respectively $\tau=\tau_+({\bm x})$, defines a regular spacelike hypersurface $\mathscr{I}^+ $.

\medskip
For part (iii) one notices that a solution of the vacuum
conformal Einstein field equations implies a solution to the vacuum
Einstein field equations ---see e.g. \cite{Fri83,Fri91}.  If
$\omega^k$ denotes the dual cobasis of the frame $e_i$, $\langle
\omega^k, e_i\rangle= \delta^k{}_i$, then the unphysical metric is
given by $g=\eta_{ij}\omega^i \otimes \omega^j$. If $\varepsilon$ is
sufficiently small, one has that $\mbox{det}(g)\neq 0$ on $[0,\tau_*]$
as $\mbox{det}(\mathring{g})\neq 0$. Thus $\tilde{g} =\Theta^{-2} g$
is well defined on $\tilde{\mathcal{M}}$.  The matter fields
$\tilde{\rho}$ and $\tilde{u}^\mu$ are defined via the formulae in
\eqref{ConformalTransformation:MatterFields}. An adaptation of this
argument to our setting gives the desired result. Geodesic
completeness follows from an analysis of the geodesic equations and
standard perturbative arguments for ordinary equations given that the
background solution is future geodesically complete.

\end{proof}

\section*{Acknowledgements}
Part of this research was carried out at the Erwin Schr\"odinger
International Institute for Mathematical Physics of the University of Vienna, Austria, during the course of
the programme ``Dynamics of General Relativity: Numerical and
Analytical Approaches'' (July-September, 2011) and the workshop
``Cartan connections, geometry of homogeneous spaces, and
dynamics'' (July 2011). The authors thank the organisers for the invitation to
attend these programmes and the institute for its hospitality. We
have profited from interesting discussions with
Prof. H. Friedrich. C.L. would like to thank Queen Mary University of London for a Visiting Fellowship.


\begin{thebibliography}{10}

\bibitem{AngTod99a}
K.~Anguige \& K.~P. Tod,
\newblock {\em Isotropic Cosmological Singularities I. Polytropic Perfect Fluid
  Spacetimes},
\newblock {A}nn. {P}hys. {\bf 276}, 257 (1999).

\bibitem{BicSchTod10b}
J.~Bi\v{c}\'{a}k, M.~Scholtz, \& K.~P. Tod,
\newblock {\em On asymptotically flat solutions of Einstein's equations
  periodic in time: II. Spacetimes with scalar-field sources},
\newblock Class. Quantum Grav. {\bf 27}, 175011 (2010).

\bibitem{Cho08}
Y.~Choquet-Bruhat,
\newblock {\em General Relativity and the Einstein equations},
\newblock Oxford University Press, 2008.

\bibitem{Fri81b}
H.~Friedrich,
\newblock {\em The asymptotic characteristic initial value problem for
  {Einstein}'s vacuum field equations as an initial value problem for a
  first-order quasilinear symmetric hyperbolic system},
\newblock Proc. Roy. Soc. Lond. A {\bf 378}, 401 (1981).

\bibitem{Fri81a}
H.~Friedrich,
\newblock {\em On the regular and the asymptotic characteristic initial value
  problem for {Einstein}'s vacuum field equations},
\newblock Proc. Roy. Soc. Lond. A {\bf 375}, 169 (1981).

\bibitem{Fri83}
H.~Friedrich,
\newblock {\em Cauchy problems for the conformal vacuum field equations in
  General Relativity},
\newblock Comm. Math. Phys. {\bf 91}, 445 (1983).

\bibitem{Fri85}
H.~Friedrich,
\newblock {\em On the hyperbolicity of Einstein's and other gauge field
  equations},
\newblock Comm. Math. Phys. {\bf 100}, 525 (1985).

\bibitem{Fri86a}
H.~Friedrich,
\newblock {\em On purely radiative space-times},
\newblock Comm. Math. Phys. {\bf 103}, 35 (1986).

\bibitem{Fri86b}
H.~Friedrich,
\newblock {\em On the existence of n-geodesically complete or future complete
  solutions of {E}instein's field equations with smooth asymptotic structure},
\newblock Comm. Math. Phys. {\bf 107}, 587 (1986).

\bibitem{Fri91}
H.~Friedrich,
\newblock {\em On the global existence and the asymptotic behaviour of
  solutions to the Einstein-Maxwell-Yang-Mills equations},
\newblock J. Diff. Geom. {\bf 34}, 275 (1991).

\bibitem{Fri95}
H.~Friedrich,
\newblock {\em {Einstein} equations and conformal structure: existence of
  anti-de {Sitter}-type space-times},
\newblock J. Geom. Phys. {\bf 17}, 125 (1995).

\bibitem{Fri98c}
H.~Friedrich,
\newblock {\em Evolution equations for gravitating ideal fluid bodies in
  general relativity},
\newblock Phys. Rev. D {\bf 57}, 2317 (1998).

\bibitem{Fri03a}
H.~Friedrich,
\newblock {\em Conformal Einstein evolution},
\newblock in {\em The conformal structure of spacetime: Geometry, Analysis,
  Numerics}, edited by J.~Frauendiener \& H.~Friedrich, Lecture Notes in
  Physics, page~1, Springer, 2002.

\bibitem{GriPod09}
J.~B. Griffiths \& J.~Podolsk\'y,
\newblock {\em Exact space-times in Einstein's General Relativity},
\newblock Cambridge University Press, 2009.

\bibitem{Hub95}
P.~H\"{u}bner,
\newblock {\em General relativistic scalar-field models and asymptotic
  flatness},
\newblock Class. Quantum Grav. {\bf 12}, 791 (1995).

\bibitem{Kat75}
T.~Kato,
\newblock {\em The Cauchy problem for quasi-linear symmetric hyperbolic
  systems},
\newblock Arch. Ration. Mech. Anal. {\bf 58}, 181 (1975).

\bibitem{KreLor98}
H.-O. Kreiss \& J.~Lorenz,
\newblock {\em Stability for time-dependent differential equations},
\newblock Acta Numerica {\bf 7}(203) (1998).

\bibitem{LueVal09a}
C.~L\"ubbe \& J.~A. {Valiente Kroon},
\newblock {\em On de Sitter-like and Minkowski-like spacetimes},
\newblock Class. Quantum Grav. {\bf 26}, 145012 (2009).

\bibitem{LueVal10}
C.~L\"ubbe \& J.~A. {Valiente Kroon},
\newblock {\em A stability result for purely radiative spacetimes},
\newblock J. Hyp. Diff. Eqns. {\bf 7}, 545 (2010).

\bibitem{LueVal11}
C.~L\"ubbe \& J.~A. {Valiente Kroon},
\newblock {\em The extended Conformal Einstein field equations with matter: the
  Einstein-Maxwell system}, 2011.

\bibitem{PenRin84}
R.~Penrose \& W.~Rindler,
\newblock {\em Spinors and space-time. {V}olume 1. {T}wo-spinor calculus and
  relativistic fields},
\newblock Cambridge University Press, 1984.

\bibitem{Ren92b}
A.~D. Rendall,
\newblock {\em The initial value problem for a class of general relativistic
  fluid bodies},
\newblock J. Math. Phys. {\bf 33}, 1047 (1992).

\bibitem{Reu99}
O.~Reula,
\newblock {\em Exponential decay for small nonlinear perturbations of expanding
  flat homogeneous cosmologies},
\newblock Phys. Rev. D {\bf 60}, 083507 (1999).

\bibitem{RodSpe09}
I.~Rodnianski \& J.~Speck,
\newblock {\em The Stability of the Irrotational Euler-Einstein System with a
  Positive Cosmological Constant},
\newblock In {\tt arXiv:0911.5501}, 2009.

\bibitem{Som80}
P.~Sommers,
\newblock {\em Space spinors},
\newblock J. Math. Phys. {\bf 21}, 2567 (1980).

\bibitem{Spe11}
J.~Speck,
\newblock {\em The Nonlinear Future-Stability of the FLRW Family of Solutions
  to the Euler-Einstein System with a Positive Cosmological Constant},
\newblock In {\tt arXiv:1102.1501}, 2011.

\end{thebibliography}

\end{document}